\documentclass[journal]{IEEEtran}

\usepackage{generic}
\usepackage{cite}
\usepackage{amsmath,amssymb,amsfonts}
\usepackage{graphicx}
\usepackage{textcomp}

\usepackage{epsfig} 
\usepackage{subfigure}

\usepackage[hidelinks]{hyperref}

\allowdisplaybreaks

\usepackage{listings}
\usepackage{algorithm}

\usepackage{color}
\usepackage{amssymb}
\usepackage{extarrows}

\newtheorem{theorem}{Theorem}

\newtheorem{remark}{Remark}
\newcommand{\lfteqn}{\begin{eqnarray} \begin{array}{llllllllll}}
\newcommand{\ndeqn}{\end{array} \nonumber \end{eqnarray}}
\newcommand{\Lfteqn}{\begin{eqnarray} \begin{array}{llllllllll}}
\newcommand{\Ndeqn}{\end{array}  \end{eqnarray}}
\def\bull{\vrule height 1.1ex width 1.1ex depth -.0ex }

\newcommand{\cS}{\mathcal{S}}
\newcommand{\cP}{\mathcal{P}}
\newcommand{\cH}{\mathcal{H}}

\def\BibTeX{{\rm B\kern-.05em{\sc i\kern-.025em b}\kern-.08em T\kern-.1667em\lower.7ex \hbox{E}\kern-.125emX}}

\begin{document}

\title {Modular Control of Discrete Event System for Modeling and Mitigating Power System Cascading Failures}

\author{Wasseem Al-Rousan, \textit{Member, IEEE}, Caisheng Wang, \textit{Fellow, IEEE}, and Feng Lin, \textit{Fellow, IEEE} 
	\vspace{-2em}
	
	\thanks{Wasseem Al-Rousan is with the Department of Electrical Engineering, Philadelphia University, Amman, Jordan. Caisheng Wang and Feng Lin are with the Department of Electrical and Computer Engineering, Wayne State University, Detroit, MI 48202, USA (e-mail: walrousan@philadelphia.edu.jo; cwang@wayne.edu; flin@wayne.edu). The work of C. Wang and F. Lin on this paper was partially supported by the National Science Foundation of the USA under Grant ECCS-2146615.} }

\maketitle

\begin{abstract}
	
	Cascading failures in power systems, triggered by the sequential tripping of system components, pose a serious threat, as they can result in partial or complete system shutdowns, disrupting critical services and potentially causing significant economic damage and life-threatening consequences. 
	In prior work, we developed a new approach for identifying and preventing cascading failures in power systems.
	The approach uses supervisory control technique of discrete event systems (DES) by incorporating both on-line lookahead control and forcible events. 
	In this paper, we use modular supervisory control of DES to reduce computation complexity and increase the robustness and reliability of control. Modular supervisory control allows us to predict and mitigate cascading failures in power systems effectively. 
	We implemented the proposed control technique on a simulation platform developed in MATLAB and applied the proposed DES controller. 
	The calculations of modular supervisory control of DES are performed using an external tool and imported into the MATLAB platform. 
	We conduct simulation studies for the IEEE 30-bus, 118-bus and 300-bus systems, and the results demonstrate the effectiveness of our proposed approach.

\end{abstract}


\section*{Nomenclature}

\addcontentsline{toc}{section}{Nomenclature}
\begin{IEEEdescription}[\IEEEusemathlabelsep\IEEEsetlabelwidth{$V_1,V_2,V_3$}]
	\item[$\cP$] Plant automaton.
	\item[$K$] Safe/legal/desired language, also called the specification.
	\item[$\cS$] Supervisor/controller.
	\item[$\cH$] Automaton that generates language $K$, that is, $K = L(\cH)$. $\cH \sqsubseteq \cP$.
	\item[$K^{\uparrow}$] Supremal F-controllable sublanguage of $K$.
	\item[$\cH^{\uparrow}$] Automaton that generates language $K^{\uparrow}$.
	\item[$\cS^{\uparrow \diamond}$] Supervisor/controller that achieves $K^{\uparrow}$.
	\item[$L(\cS^{\uparrow \diamond}/\cP)$] The language generated by the supervised system.
	\item[$\cP_{i}$] Automaton of component $i$.
	\item[$\cP^{j}$] Automaton of sub-system $j$.
	\item[$K_j$] Safe/legal/desired language of sub-system $j$.
	\item[$\cH_j$] Automaton that generates language $K_j$.
	\item[$K^{\uparrow}_{j}$] Supremal F-controllable sublanguage of $K_j$.
	\item[$\cH^{\uparrow}_{j}$] Automaton that generates language $K^{\uparrow}_{j}$.
	\item[$\cS^{\uparrow \diamond}_{j}$] Supervisor/controller that achieves $K^{\uparrow}_{j}$.
	\item[$L(\cS^{\uparrow \diamond}_j/\cP^j)$] The language generated by the supervised sub-system $j$.
	\item[$L(\wedge \cS^{\uparrow \diamond}_j/\cP)$] The language generated by the supervised system with modular controllers.		
	\item[$\delta$] The partial transition function, $\delta(q,\sigma) = q'$ means $q \xrightarrow \sigma q'$. $\delta$ is described by the state transition diagram of the automaton.
	\item[$\delta_H$] The partial transition function for the Automaton $\cH$.
	\item[$\delta_{H}^{\uparrow}$] The partial transition function for the Automaton $\cH^{\uparrow}$.
	\item[$\delta_{H}^{j}$] The partial transition function for the Automaton $\cH^{j}$ for sub-system $j$.		
	\item[$\delta_{H}^{\uparrow j}$] The partial transition function for the Automaton $\cH^{\uparrow j}$ for sub-system $j$ .
	\item[$\Sigma$] Events set.
	\item[$\Sigma_c$] Controllable events set.
	\item[$\Sigma_{uc}$] Uncontrollable events set.
	\item[$\Sigma_f$] Forcible events set.		
	\item[$\Sigma_{uf}$] Unforcible events set.		
\end{IEEEdescription}

\section{Introduction}
\label{sec:introduction}

Cascading failures in power systems are a significant concern due to the interconnected nature of electrical grids and the potential for a localized issue to propagate and cause widespread outages. 
The causes of cascading failures include (1) overloading of transmission lines or transformers, (2) voltage instability, (3) equipment failures, (4) natural disasters, (5) cyberattacks, and more.

Because they can lead to widespread blackouts, disrupt vital services and cause significant economic and social losses, cascading failures have been investigated extensively in the literature from different points of view. Among the papers published, the ones using abstracted and higher-level models are most relevant to our work as described below.  Influence graphs are used to investigate the probability functions of the initial outages of branches and the outages that occur consequently in \cite{hines2016cascading}. 
In \cite{Bialek7404289}, the authors surveyed the tools used to analyze and simulate cascading failures, and provided recommendations for benchmarking and validation among tools and methods. Additionally, test cases used in the analysis and benchmarking were discussed. In \cite{Li10497182}, the authors investigated cascading failure topological properties, where the electrical distance between consecutive failures was analyzed. Power system evolution during cascading failure was measured. 
In \cite{asavathiratham2001influence}, transition probabilities between states in automata representing components of power systems were formalized. Information about the most vulnerable lines using Markovian models was analyzed in \cite{AMarkovianInfluenceGraph} based on line outage data. 
In \cite{Cutset1} and \cite{Cutset2}, transmission line failure propagation is characterized using a network graph structure. Two cases are considered: when the network remains connected after a contingency and when the failure propagation separates the network into multiple islands. Other graph theory based methods are proposed in \cite{SPERSTAD2021106408, YANG2021107354, Resilienceassessment, Ahybridapproachtransmission}, \cite{Athresholdmodel} for studying cascading failures. \cite{LIU20227908} presents a method to identify vulnerable branches by considering the propagation characteristics of cascading failures. Network failures initiated by link failures and overloads are studied in \cite{PerezPhysRevE}. A tree partitioning method based on the assumption that a given power system can be represented as a connection of clusters is proposed to mitigate cascading failure \cite{Tree_Partitioning}. A distributed model based predictive control to mitigate cascading failure risk is proposed in\cite{Reciprocally}.

Many approaches have been proposed to analyze, model, and control power systems under various  scenarios. One approach that considers reducing communication burden, increasing reliability, improving efficiency, and reducing computational load is by using an event-based approach to monitor and control the power system. In \cite{Abdelmalak9893118}, the authors performed a survey on modeling methods of cyber-physical power systems. One of the modeling approaches was the use of finite-state machines and event-based models. In \cite{Zhu10561600}, the authors studied peer-to-peer transactions to enhance distribution system resilience in case of high-impact, low-probability events that may lead to cascading failures, where distributed energy resources participate in the resilience enhancement.
The authors in \cite{Bu9626584} proposed an event-triggered load frequency control based on model-free adaptive control, which depends on the system data. In \cite{Yang9737119}, the authors proposed a dynamical event-triggered controller that reduces the usage of communication resources to control multi-area wind power systems under dual alterable aperiodic denial-of-service attacks. The authors in \cite{ghorashi2021distributed} proposed a rule-based supervisory control system to avoid the state of charge violations in a hybrid energy storage system. The authors in \cite{Liu9756376} proposed a hybrid event-triggered control approach that has both continuous and discrete dynamics to describe the dynamics of power systems under denial of service attacks. In \cite{Abdelmalak9693217}, the authors proposed a probabilistic proactive strategy to enhance power system resilience against wildfires, formulated within a Markov decision process framework, wherein the system is modeled as a set of states with transitions governed by event occurrences.

Cascading failures in power systems can be viewed as a string of events leading to widespread disruptions or outages. To investigate cascading failures, it is natural to model a system as a discrete event system (DES) at some level of abstraction. 
Controllers in DES are also called supervisors (we use controller and supervisor interchangeably) and control is called supervisory control \cite{ramadge1987supervisory, lin1988observability, WONHAMbriefhistory}. We have proposed a framework using DES and supervisory control to model and mitigate cascading failure processes in \cite{ADiscretecascading, LLPWasseem, thesisWasseem}.
The framework is as follows. To construct a DES model for a power system, we first model its components as automata. Since these automata are small, they can be easily obtained. 
We then combine these automata using parallel composition to obtain the automaton for the entire power system. Because the parallel composition can be performed by existing DES software, it can be done automatically by computers.

Since some events, such as line trips, cannot be disabled but can be preempted by forcing some forcible events, such as load shedding, the conventional supervisory control of DES is not adequate to deal with cascading failures in power systems. 
We extend supervisory control of DES to include forcible events in \cite{ADiscretecascading, LLPWasseem, thesisWasseem}, which not only provides a solution to the cascading failure problems, but also significantly increases the applicability of supervisory control.
To manage computational complexity, we propose an online lookahead control, which significantly reduces the number of states to be considered.
We also implement the proposed online lookahead control in an implementation platform using MATLAB.  
The platform uses MATPOWER to simulate a power system and then control it using the proposed DES controller. Simulation studies are carried out for IEEE 6-, 30-, and 118-bus systems. The results verify the effectiveness of the previously proposed framework.

To control large power systems, the DES-based centralized control proposed in \cite{ADiscretecascading, LLPWasseem, thesisWasseem} may not be adequate, because of large number of states involved. 
A modular/decentralized control that divides the system and control task into several sub-systems and sub-tasks, each achieved by a local controller, is more appropriate for large power systems for the following reasons. (1) Preventing cascading failures is a time-critical task. Due to large number of states involved, it may take a centralized controller too long to calculate the needed control. 
(2) A centralized controller may not be robust to communication delays and losses, which are unavoidable. Hence, it is better to have a set of local and decentralized controllers with much smaller communication delays and losses to prevent cascading failures. 

Modular supervisory control of DES without forcible events has been investigated in the literature \cite{wonham1988modular, wong1998modular, komenda2007control, zhou2007decentralized}. 
They have also been applied to manufacturing systems \cite{s23010163}, freeway traffic control \cite{Pasquale9178965}, and multiple UAVs \cite{KarimoddiniDecentralized}. 
These applications show that modular supervisory control can make a significant difference in terms of computational complexity and robustness.

To use modular control for preventing cascading failures, we first need to extend conventional modular supervisory control to allow forcible events. Therefore, in this paper, we formally define a modular supervisory control mechanism with forcible events. 
We then investigate the behavior (language) generated by the modular supervised system. We derive a necessary and sufficient condition for the existence of modular controllers, which is F-controllability and conditional decomposability. 
If the conditions are satisfied, we will design the modular controllers. If the conditions are not satisfied, then we will design the modular controllers that generate a smaller language and hence ensure no cascading failures in the controlled system, while, at the same time, giving the controlled system maximum freedom to achieve other control objectives.
\par DES-based supervisory controllers have been implemented in hardware setups in previous attempts, such as the work in \cite{ghasaei2020discrete}, where a supervisory control strategy was implemented in a Programmable Logic Controller (PLC) and tested in a Hardware in the Loop (HIL) setup. Several applications of DES-based control in power systems can be found in \cite{ROMERORODRIGUEZ201997}, \cite{ozbaltan2025control}, \cite{kharrazi2017discrete}, and \cite{carati2021supervisory}. The proposed framework can be implemented using a separate controller that may be located at substations (nodes), with access to communication resources to substations that are directly connected by transmission lines. This controller has the ability to perform load shedding and generation re-dispatch.

After developing the theoretical framework for modular supervisory control with forcible events, we implement the results in large scale power systems. The implementation combines two simulation platforms based on \cite{dcsimsep} and \cite{libFAUDES} to model and mitigate cascading failures in power systems. The proposed approach was compared with several existing methods, as shown in Table \ref{Table_4}, with respect to the general procedures employed in cascading failure mitigation and the associated evaluation metrics.

\begin{table*}[h!]

	\caption{Comparison among different methods}
	\vspace{-20pt}
	\label{Table_4}
	\footnotesize
	\begin{center}
		\begin{tabular}{|l l l l |} 
			
			\hline
			Method & Ref. & Description & Evaluation metrics\\ [0.5ex] 
			\hline\hline
			On-Line DES &	\cite{LLPWasseem}  &  Centralized DES-based limited lookahead control  & Expected MW lost (proportional to expected blackout size)  \\
			\hline
			Modular control DES &	  &  Modular DES supervisory controllers & $1-$  Expected MW lost (proportional to expected blackout size)  \\
			&	  &   & $2-$ Complementary cumulative distribution (CCD)  \\
			&	  &   & of line outages and blackout size \\
			\hline
			Emergency control &	\cite{thesis}  &  Linear programming problem to minimize  & Expected blackout size  \\
			&	  &  load lost within power flow constraints  &   \\
			\hline
			Decentralized MPC &	\cite{HinesDecMPC} \cite{Reciprocally} &  Decentralized model predictive control & Comparison of blackout size when changing the size of local\\
			&	  &  where agents act with reciprocal altruism & neighborhood of control agents \\
			\hline
			Influence graph & \cite{hines2016cascading}  &  A Markovian influence graph & Critical components metric $\alpha_j$. And expected blackout size \\
			&   &  that is constructed from cascading failure simulations & before and after increasing power flow limits  \\
			&   &  & for most critical components
			\\
			\hline
			Markovian tree & \cite{kaisun} \cite{ManagementofCascading} &  Markovian tree risk based management & Risk cost (MW lost) versus control cost  \\
			&   &   based on risk gradient &  (generation adjusting) converted to economic metrics  \\
			\hline
		\end{tabular}
		
	\end{center}
\end{table*}

The main innovations and contributions of the paper are as follows. (1) We extend the theory of supervisory control with forcible events from centralized control to modular (decentralized) control. This extension significantly reduces the computational complexity and increases robustness and reliability. 
(2) Using this extension, we propose a new control method to prevent cascading failures in large scale power. This new method uses modular controllers that make control decisions based on the information received from neighboring nodes and send control actions to the local nodes, making the control faster and more reliable. 
(3) We develop a new platform based on MATLAB environment that combines discrete events and continuous variables to implement the proposed modular control. 
(4) Using the platform, we implement and simulate modular control for the IEEE 30-bus, 118-bus, and 300-bus systems. The results show the effectiveness of the proposed method.

The remainder of the paper is organized as follows. Section II introduces DES and necessary notations. Section III discusses supervisory control with forcible events. Section IV presents the modular supervisory control theory with forcible events. Section V applies the results to large scale power systems to prevent cascading failures by implementing the modular controllers using the proposed implementation platform. Section V also presents simulation results. The conclusion is drawn in Section VI.

\section{Discrete Event System Model of Power Systems}

A power system to be controlled, called plant, can be modeled as a discrete event system using the following automaton model:
$$
\cP=( Q, \Sigma , \delta , q_o ),
$$
where $Q$ is the set of states; $\Sigma$ is the set of events;
$\delta$ is the (partial) transition function; and $q_o$ is the initial state. Examples of automata for power line, generator, and load are shown in Section V (Fig. \ref{Basic_Model_Figure}).

An automaton can be thought of as a system model with states that represent the operational modes of a given system and events that represent the instantaneous transitions from one state to another. 
Starting from the initial state $q_o$, the system continuously moves from one state to another, generating a string of events. Hence, a trajectory of the system is represented by the corresponding string. 
The automaton model describes the set of all possible trajectories or strings, which is called the language generated by the automaton\footnote{The terminologies come from automata and formal languages theory, where alphabets represent events, sentences represent strings and languages represent the set of strings.  }.

Formally, the transition function $\delta : Q \times \Sigma \rightarrow Q$ describes the ``dynamics'' of the system in the sense that if the current state is $q \in Q$ and event $\sigma \in \Sigma$ is defined in $q$, then after the occurrence of $\sigma$, the next state is $\delta (q,\sigma)$. $\delta$ can be extended from events to strings as $\delta : Q \times \Sigma ^* \rightarrow Q$, where $\Sigma ^*$ denotes the set of all strings over $\Sigma$. It is possible that not all events can occur in a state $q$ (for example, a switch cannot be turned on again if it is already on). Hence, $\delta$ is a partial function. The language {\em generated} by $\cP$ is the set of all strings defined in the automaton $\cP$ starting at the initial state and is denoted by
\lfteqn
L(\cP) = \{ s \in \Sigma ^*: \delta (q_o,s)! \}, 
\ndeqn
where $\delta (q_o,s)!$ is used to denote that $\delta (q_o,s)$ is defined.

To obtain an automaton model for a power system, we use a modular approach by first developing (simple) models for its components as
$$
\cP _i =( Q_i, \Sigma _i , \delta _i, q_{o,i} ), \ i=1,2,...,C.
$$

In this paper, we consider transmission lines, generators and loads as the basic components of the power system. See Section V for details.
If a power system has $m$ transmission lines, $n$ generators, and $k$ loads, the power system will have $C= m+n+k$ components. The overall power system can be constructed using parallel compositions \cite{cassandras2009introduction}:
\lfteqn
\cP & = \cP _1 || \cP _2 || ... || \cP_{C}
\ndeqn

The number of states in $\cP$ can increase exponentially with the number of components. In other words, if each component has 2 states, then $\cP$ will have $2^C = 2^{m+n+k}$ states. 
To overcome this state explosion, on-line lookahead control and/or modular control can be used to significantly reduce the number of states to be considered. We investigated on-line lookahead control in \cite{LLPWasseem}. 
In this paper, we investigate modular on-line lookahead control with forcible events.

The computational complexity of control is proportional to the number of states in $\cP$. For centralized control, this number is $2^{m+n+k}$. For modular control, since each controller controls a small portion of $\cP$, this number is $2^{m_j+n_j+k_j}$, where $m_j$, $n_j$, $k_j$ are the numbers of transmission lines, generators, and loads controlled by controller $j$, respectively. Clearly, $m_j+n_j+k_j$ is much smaller than $m+n+k$. Therefore, modular control can significantly reduce the computational complexity.

\section{Supervisory Control with Forcible Events}

In control of DES, controllers are also called supervisors (we use controller and supervisor interchangeably) and the control is called supervisory control. In conventional supervisory control, a supervisor can only enable and disable events, it cannot force events.
In \cite{LLPWasseem}, we extended the conventional supervisory control of DES to allow forcible events since disablement and enablement are not sufficient in our application in power systems. In this section, we briefly review the findings. The controller proposed in \cite{LLPWasseem} can force events in addition to disabling and enabling events to achieve the control objectives. The following assumptions are made on the events.

\begin{enumerate}
	
	\item There is a set of controllable events $\Sigma _c \subseteq \Sigma$; that is, their occurrences can be disabled. The set of uncontrollable events is denoted by $\Sigma_{uc} = \Sigma - \Sigma _c$.

	\item There is a set of forcible events $\Sigma _f \subseteq \Sigma$; that is, a controller can force them to occur. We assume that forcible events can preempt uncontrollable events.
	
\end{enumerate}

In our application of preventing cascading failures in power systems, overloading of transmission lines are uncontrollable. Load shedding is a forcible events. So, for example, transmission line overloading can be prevented by load shedding.

Formally, a controller is a mapping: 
\lfteqn
\cS: L(\cP) \rightarrow 2^\Sigma.
\ndeqn
where $2^\Sigma$ denotes the set of all subsets of $\Sigma$. Hence, after the occurrence of a string $s \in L(\cP)$, the events in $\cS (s)$ can occur next.

We use $L(\cS/\cP)$ to denote the language generated by the controlled system, which is defined recursively as
\Lfteqn \label{Equation1}
(1) & \varepsilon \in L(\cS/\cP), \\
(2) & (\forall s \in L(\cS/\cP)) (\forall \sigma \in \Sigma) s
\sigma \in L(\cS/\cP) \\
& \Leftrightarrow (s \sigma \in L(\cP) \wedge \sigma \in \cS
(s)) .
\Ndeqn

After the occurrence of any string $s \in L(\cP)$, the control $\cS(s)$ must satisfy one of the following two conditions: 

\begin{enumerate}
	
	\item all events disabled are controllable, that is, $\Gamma(s) - \cS
	(s) \subseteq \Sigma _c$, where $\Gamma(s) = \{ \sigma \in
	\Sigma: s \sigma \in L(\cP) \}$; or 
	
	\item some events enabled are forcible, that is, $\cS (s) \cap \Sigma _f \cap \Gamma (s)
	\not= \emptyset$. This is based on the assumption that forcible events can preempt uncontrollable events. 
	
\end{enumerate}

Therefore, the following condition is required:
\Lfteqn \label{Equation2}
(\forall s \in L(\cS/\cP)) (\Gamma(s) - \cS (s)
\subseteq \Sigma _c ) \\
\vee (\cS (s) \cap \Sigma _f \cap \Gamma (s) \not=
\emptyset).
\Ndeqn

Let $K \subseteq L(\cP)$ be a specification language that excludes all strings leading to cascading failures in power systems. Our goal is to design a controller $\cS$ such that $L(\cS/\cP) = K$ if possible, and $L(\cS/\cP) \subseteq K$, otherwise. Note that $L(\cS/\cP) \subseteq K$ means that the controlled system will not generate any strings not in $K$, that is, there will be no cascading failures.

To derive a necessary and sufficient condition for the existence of a controller $\cS$ such that $L(\cS/\cP) = K$, F-controllability is introduced in \cite{LLPWasseem}. F-controllability extends controllability by allowing forcible events to preempt uncontrollable events and hence effectively disable the preempted uncontrollable events. In power systems, for example, line tripping is uncontrollable. However, it can be preempted by a load shedding. Formally, A language $K \subseteq L(\cP)$ is F-controllable with respect to $L(\cP)$, $\Sigma _c$, and $\Sigma _f$ if
\lfteqn
(\forall s \in K) (\forall \sigma \in \Sigma) (s \sigma
\in L(\cP) \wedge s \sigma \not\in K) \\
\Rightarrow (\sigma \in \Sigma _c \vee (\exists \sigma _f \in
\Sigma _f) s \sigma _f \in K).
\ndeqn
F-controllability says that if $\sigma$ is not allowed
after $s$, then either $\sigma$ is controllable (so that it can be
disabled) or there is another forcible event $\sigma _f$ that is
allowed and can preempt $\sigma$. If $\Sigma _f = \emptyset$, then F-controllability reduces to controllability \cite{ramadge1987supervisory}.

The following result is then obtained in \cite{LLPWasseem}: There  exists a controller $\cS: L(\cP) \rightarrow 2^\Sigma$ such that $L(\cS/\cP)=K$ if and only if $K$ is F-controllable (with respect to $L(\cP)$, $\Sigma _c$, and $\Sigma _f$) \footnote{\cite{LLPWasseem} considers partial observation (that is, not all events are observable). Therefore, observability \cite{lin1988observability} is also required. Since we consider full observation in this paper, observability is not required.}.

The specification language $K$ is often generated by a sub-automaton $\cH \sqsubseteq \cP$, that is, $K = L(\cH)$ for some
$$
\cH=( Q_H, \Sigma , \delta_H , q_o ),
$$
where $Q_H \subseteq Q$ and $\delta_H = \delta |_{Q_H \times \Sigma} \subseteq \delta$ ($\delta |_{Q_H \times \Sigma}$ means $\delta$ restricted to $Q_H$). In this way, the state set $Q$ is partitioned into legal/safe state set $Q_H$ and illegal/unsafe state set $Q-Q_H$. 

If $K$ is F-controllable, then the following controller $\cS^\diamond$ achieves $K$, that is, $L(\cS^\diamond/\cP)=K$. 
\Lfteqn \label{Equation3}
\cS^\diamond (s) = \{\sigma \in \Sigma : s \sigma \in K \}.
\Ndeqn

If $K$ is not F-controllable, then we would like to find a sublanguage of $K' \subseteq K$ that is F-controllable and to design a controller $\cS$ such that $L(\cS/\cP)=K'$. 
To give the controlled system maximum freedom to perform other tasks, we would like to make $K'$ as large as possible. In other words, we would like to find the supremal (largest) F-controllable sublanguage of $K$, denoted by $K^\uparrow$. 
By our result in \cite{LLPWasseem}, $K^\uparrow$ exists and is unique, because the union of F-controllable languages is also F-controllable.

$K^\uparrow$ can be obtained by iteratively removing ``bad'' states in $\cH$. A state $q \in Q_H$ is bad if
\lfteqn
(\exists \sigma \in \Sigma _{uc}) \delta (q,\sigma) \in Q \wedge
\delta (q,\sigma) \not\in Q_H \\
\wedge (\forall \sigma _f \in \Sigma _f) \delta (q,\sigma_f)
\not\in Q_H).
\ndeqn
Denote the automaton after removing all bad states as
$$
\cH^\uparrow=( Q_H^\uparrow, \Sigma , \delta_H^\uparrow , q_o ),
$$
where $\delta_H^\uparrow = \delta_H|_{Q_H^\uparrow \times \Sigma}$. Then $K^\uparrow = L(\cH^\uparrow)$. Clearly, $Q_H^\uparrow \subseteq Q_H$ and $K^\uparrow \subseteq K$. Note that if $K$ is F-controllable, then $K^\uparrow = K$ and $\cH^\uparrow = \cH$. The controller $\cS^{\uparrow \diamond}$ achieves $K^\uparrow$ is given by 
\Lfteqn \label{Equation4}
\cS^{\uparrow \diamond} (s) = \{\sigma \in \Sigma : s \sigma \in K^\uparrow \}.
\Ndeqn

Since $L(\cS^{\uparrow \diamond}/\cP)=K^\uparrow \subseteq K$, $\cS^{\uparrow \diamond}$ ensures that the controlled system stays within the legal specification language $K$; that is, it can prevent cascading failures. 
Furthermore, since $K^\uparrow$ is the supremal F-controllable sublanguage of $K$, $\cS^{\uparrow \diamond}$ gives the system maximum freedom without violating legal specification, that is, it will not disable/preempt an event unless it is absolutely necessary to do so (not disabling/preempting it will lead to cascading failures).

Constructing $\cS^{\uparrow \diamond}$ off-line requires that we construct $\cP$ first. Since $\cP$ is the parallel composition of all components $\cP_{i}$, the number of states in $\cP$ can grow exponentially with respect to the number of components. 
To overcome this state explosion problem, we propose on-line control using limited lookahead policies in \cite{LLPWasseem}. For on-line control, we construct a lookahead tree until the number of steps/levels reaches $M$, the limit on lookahead steps.
%
The limited lookahead tree from the current state $q \in Q$ is denoted by
$$
Tree(q) = (Y, \Sigma, \zeta, y_o).
$$

In \cite{LLPWasseem}, a controller based on $Tree(q)$ with a limited lookahead policy of $M$ steps, denoted by $\cS ^M_{CLL}$, is proposed such that $L(\cS ^M_{CLL}/\cP) \subseteq K$. Simulation studies are also carried out for IEEE 6-bus, 30-bus, and 118-bus systems in \cite{LLPWasseem}, which verify the effectiveness of the proposed $\cS ^M_{CLL}$.

\section{ Modular supervisory control with forcible events}

For large scale power systems, the centralized control proposed in \cite{LLPWasseem} may not be effective for the following reasons. 
(1) The complexity of the lookahead tree and hence the time needed to compute the control increases as the number of components increases. Since preventing cascading failures is a time-critical task, it may take a lookahead tree-based centralized controller too long to compute the control under the current computational capabilities, since the size of the lookahead tree increases exponentially as the system size grows. 
(2) The centralized controller requires each node in a power system to communicate its information to the central controller, which may not be the most reliable because it has a single point of failure.
(3) Communication delays and losses are much larger between a centralized controller and distributed actuators than between local controllers and actuators. 

To address the above issues, modular supervisory  control has been proposed for DES. In modular supervisory  control, the uncontrolled system and control tasks are divided into several sub-systems and sub-tasks. Control is then achieved using several modular controllers, each for a sub-system and a sub-task.

Modular supervisory control is more robust and reliable for large scale power systems. However, in order to use modular supervisory control in power systems, we need to extend modular supervisory control to allow forcible events. We will do so in this section.

For modular supervisory control, we assume that the uncontrolled system is the parallel composition of $n$ sub-systems, that is, 
\Lfteqn \label{Equation5}
\cP = \cP^{1}||\cP^{2} || ... || \cP^{n}.
\Ndeqn
Note that $\cP^j$ may itself be the parallel composition of some components $\cP_i$, that is, $\cP^j = \cP _1 || \cP _2 || ... || \cP_{C_j}$. Denote $\cP^i$ as
$$
\cP ^j =( Q^j, \Sigma ^j , \delta ^j, q_o^j ), \ j=1,2,...,n.
$$

For sub-system $\cP^j$, its local forcible events are denoted by $\Sigma _f ^j = \Sigma _f \cap \Sigma ^j$; its local controllable events are denoted by $\Sigma _c ^j = \Sigma _c \cap \Sigma ^j$; and its local uncontrollable events are denoted by $\Sigma _{uc} ^j = \Sigma _{uc} \cap \Sigma ^j$, $j=1,2,...,n$. We assume that all events are observable. Denote the projection from the global events  $\Sigma$ to local events $\Sigma ^j$ by $\theta_{j}$.
Control is achieved by $n$ modular controllers
$$
\cS _j: L(\cP ^j) \rightarrow 2^{\Sigma^j} , \ j=1,2,...,n.
$$ 
The controlled sub-systems are denoted by $\cS_j/\cP^j$.
The language generated by the controlled sub-system, denoted by $L(\cS_j/\cP^j)$, is defined recursively as
\Lfteqn \label{Equation6}
(1) & \varepsilon \in L(\cS_j/\cP^j), \\
(2) & (\forall s \in L(\cS_j/\cP^j)) (\forall \sigma \in \Sigma^j) s
\sigma \in L(\cS_j/\cP^j) \\
& \Leftrightarrow (s \sigma \in L(\cP^j) \wedge \sigma \in \cS_j
(s)) .
\Ndeqn

The overall control is the conjunction of all modular controllers, denoted by $\wedge \cS_j=\cS_1 \wedge \cS_2 \wedge ... \wedge \cS_n$, that is, for any $s \in L(\cP)$,
\Lfteqn \label{Equation7}
& \wedge \cS_j(s) \\
= & (\cS_1 (\theta_{1}(s)) \cup (\Sigma - \Sigma ^1)) \cap (\cS_2 (\theta_{2}(s)) \cup (\Sigma - \Sigma ^2)) \\
&  \cap ... \cap (\cS_n (\theta_{n}(s)) \cup (\Sigma - \Sigma ^n)) .
\Ndeqn
In other words, an event $\sigma$ is allowed ($\sigma \in \wedge \cS_j(s)$) if it is allowed by all controllers ($\sigma \in \cS_i(s)$) whose event set includes $\sigma$ ($\sigma \in \Sigma_i$).

\begin{figure} [h!]
	\centering
	\includegraphics[keepaspectratio=true,angle=0, width= 0.6 \linewidth]{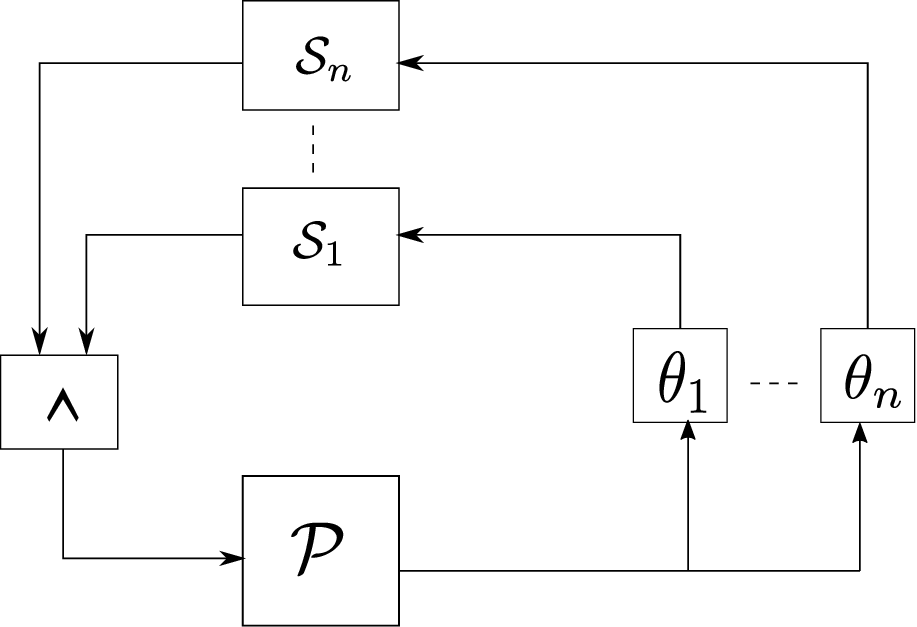}
	\caption{Modular supervisory control }
	\label{Figure_2_modular}
\end{figure}

The closed-loop system under modular control is denoted by 
$\wedge \cS_j/\cP$, which is shown in Fig. \ref{Figure_2_modular}.

Similar to centralized control, we require that, for all $j=1,2,...,n$,
\lfteqn
(\forall s \in L(\cS_j/\cP^j)) (\Gamma_j(s) - \cS_j (s)
\subseteq \Sigma _c^j ) \\
\vee (\cS_j (s) \cap \Sigma _f^j \cap \Gamma_j (s) \not=
\emptyset), 
\ndeqn
where $\Gamma_j(s) = \{ \sigma \in \Sigma^j: s \sigma \in L(\cP^j) \}$. To ensure that any event forced by $\cS _j$ is actually allowed, we further require that, for all $j=1,2,...,n$,
\Lfteqn \label{Equation8}
(\forall s \in L(\cS/\cP)) (\forall \sigma \in \Sigma) \\
\sigma \in \cS_j (\theta_{j}(s)) \cap \Sigma _f^j \cap \Gamma_j (\theta_{j}(s)) \Rightarrow s \sigma \in L(\cS/\cP) .
\Ndeqn

\begin{remark}
	Unlike centralized control, where a forcible event can only be forced by one (centralized) controller, in modular control, a forcible event may be forced by more than one (modular) controller. Equation (\ref{Equation8}) ensures that we can take the disjunction of the set of events forced by modular controllers as the set of forced events. In fact, whichever event is first forced by a modular controller will occur. 
\end{remark}

\begin{theorem}
	The language generated by the closed-loop system, $L(\wedge \cS_j/\cP)$ is given by
	\lfteqn
	L(\wedge \cS_j/\cP) = & \theta _1 ^{-1} (L(\cS_1/\cP^1)) \cap \theta _2 ^{-1} (L(\cS_2/\cP^2)) \\
	& \cap ... \cap \theta _n ^{-1} (L(\cS_n/\cP^n)).
	\ndeqn
\end{theorem}
{\em Proof}

We prove the result by induction. To do so, let us first note that, 
\lfteqn
& s \in \theta _1 ^{-1} (L(\cS_1/\cP^1)) \cap \theta _2 ^{-1} (L(\cS_2/\cP^2)) \\
& \cap ... \cap \theta _n ^{-1} (L(\cS_n/\cP^n)) \\
\Leftrightarrow
& s \in \theta _1 ^{-1} (L(\cS_1/\cP^1)) \wedge s \in \theta _2 ^{-1} (L(\cS_2/\cP^2)) \\
& \wedge ... \wedge s \in \theta _n ^{-1} (L(\cS_n/\cP^n)) \\
\Leftrightarrow
& \theta _1 (s) \in L(\cS_1/\cP^1) \wedge \theta _2 (s) \in L(\cS_2/\cP^2) \\
& \wedge ... \wedge \theta _n (s) \in L(\cS_n/\cP^n) \\
\ndeqn

Hence, we need to prove that, for all $s \in L(\cP)$, 
\lfteqn
& s \in L(\wedge \cS_j/\cP) \\
\Leftrightarrow
& \theta _1 (s) \in L(\cS_1/\cP^1) \wedge \theta _2 (s) \in L(\cS_2/\cP^2) \\
& \wedge ... \wedge \theta _n (s) \in L(\cS_n/\cP^n) .
\ndeqn
We prove this by induction of the length $|s|$ of $s$ as follows.

\noindent {\em Induction Base:} Since $\varepsilon \in L(\wedge \cS_j/\cP)$ and $\varepsilon \in L(\cS_j/\cP^j)$, for
$|s| = 0$, that is, $s=\varepsilon$, we have
\lfteqn
& s \in L(\wedge \cS_j/\cP) \\
\Leftrightarrow
& \theta _1 (s) \in L(\cS_1/\cP^1) \wedge \theta _2 (s) \in L(\cS_2/\cP^2) \\
& \wedge ... \wedge \theta _n (s) \in L(\cS_n/\cP^n) \\
\ndeqn

\noindent {\em Induction Hypothesis:} Assume that for all $s \in \Sigma^*$, $|s| \leq m$,
\lfteqn
& s \in L(\wedge \cS_j/\cP) \\
\Leftrightarrow
& \theta _1 (s) \in L(\cS_1/\cP^1) \wedge \theta _2 (s) \in L(\cS_2/\cP^2) \\
& \wedge ... \wedge \theta _n (s) \in L(\cS_n/\cP^n) \\
\ndeqn

\noindent {\em Induction Step:} We show that for all $s \in \Sigma^*$, $\sigma \in \Sigma$, $|s\sigma| = m+1$,
\lfteqn
& s\sigma \in L(\wedge \cS_j/\cP) \\
\Leftrightarrow
& \theta _1 (s\sigma) \in L(\cS_1/\cP^1) \wedge \theta _2 (s\sigma) \in L(\cS_2/\cP^2) \\
& \wedge ... \wedge \theta _n (s\sigma) \in L(\cS_n/\cP^n) \\
\ndeqn

Indeed,
\begin{align*}
	& s \sigma \in L(\wedge \cS_j/\cP) \\
	\Leftrightarrow
	& s \in L(\wedge \cS_j/\cP) \wedge s \sigma \in L(\cP) \wedge \sigma \in \wedge \cS_j (s)  \\
	& (\mbox{by Equation (\ref{Equation1})}) \\
	\Leftrightarrow
	& s \in L(\wedge \cS_j/\cP) \wedge s \sigma \in L(\cP) \\
	& \wedge \sigma \in (\cS_1 (\theta_{1}(s)) \cup (\Sigma - \Sigma ^1)) \\
	& \cap (\cS_2 (\theta_{2}(s)) \cup (\Sigma - \Sigma ^2)) \\
	&  \cap ... \cap (\cS_n (\theta_{n}(s)) \cup (\Sigma - \Sigma ^n)) \\
	& (\mbox{by Equation (\ref{Equation7})}) \\
	\Leftrightarrow
	& s \in L(\wedge \cS_j/\cP) \wedge s \sigma \in L(\cP) \\
	& \wedge \sigma \in (\cS_1 (\theta_{1}(s)) \cup (\Sigma - \Sigma ^1)) \\
	& \wedge \sigma \in (\cS_2 (\theta_{2}(s)) \cup (\Sigma - \Sigma ^2)) \\
	&  \wedge ... \wedge \sigma \in (\cS_n (\theta_{n}(s)) \cup (\Sigma - \Sigma ^n)) \\
	\Leftrightarrow
	& s \in L(\wedge \cS_j/\cP) \wedge s \sigma \in L(\cP) \\
	& \wedge (\sigma \in \cS_1 (\theta_{1}(s)) \vee \theta_{1} (\sigma) = \varepsilon ) \\
	& \wedge (\sigma \in \cS_2 (\theta_{2}(s)) \vee \theta_{2} (\sigma) = \varepsilon ) \\
	&  \wedge ... \wedge (\sigma \in \cS_n (\theta_{n}(s)) \vee \theta_{n} (\sigma) = \varepsilon ) \\
	\Leftrightarrow
	& \theta _1 (s) \in L(\cS_1/\cP^1) \wedge \theta _2 (s) \in L(\cS_2/\cP^2) \\
	& \wedge ... \wedge \theta _n (s) \in L(\cS_n/\cP^n) \wedge s \sigma \in L(\cP) \\
	& \wedge (\sigma \in \cS_1 (\theta_{1}(s)) \vee \theta_{1} (\sigma) = \varepsilon ) \\
	& \wedge (\sigma \in \cS_2 (\theta_{2}(s)) \vee \theta_{2} (\sigma) = \varepsilon ) \\
	&  \wedge ... \wedge (\sigma \in \cS_n (\theta_{n}(s)) \vee \theta_{n} (\sigma) = \varepsilon ) \\
	& (\mbox{by Induction Hypothesis }) \\
	\Leftrightarrow
	& \theta _1 (s) \in L(\cS_1/\cP^1) \wedge \theta _2 (s) \in L(\cS_2/\cP^2) \\
	& \wedge ... \wedge \theta _n (s) \in L(\cS_n/\cP^n) \\
	& \wedge \theta _1 (s \sigma) \in L(\cP^1) \wedge \theta _1 (s \sigma) \in L(\cP^1) \\
	& \wedge ... \wedge \theta _1 (s \sigma) \in L(\cP^1) \\
	& \wedge (\sigma \in \cS_1 (\theta_{1}(s)) \vee \theta_{1} (\sigma) = \varepsilon ) \\
	& \wedge (\sigma \in \cS_2 (\theta_{2}(s)) \vee \theta_{2} (\sigma) = \varepsilon ) \\
	&  \wedge ... \wedge (\sigma \in \cS_n (\theta_{n}(s)) \vee \theta_{n} (\sigma) = \varepsilon ) \\
	& (\mbox{by Equation (\ref{Equation5})}) \\
	\Leftrightarrow
	& \theta _1 (s) \in L(\cS_1/\cP^1) \wedge \theta _1 (s) \theta _1 (\sigma) \in L(\cP^1) \\
	& \wedge (\sigma \in \cS_1 (\theta_{1}(s)) \vee \theta_{1} (\sigma) = \varepsilon ) \\
	& \wedge \theta _2 (s) \in L(\cS_2/\cP^1) \wedge \theta _2 (s) \theta _2 (\sigma) \in L(\cP^2) \\
	& \wedge (\sigma \in \cS_2 (\theta_{2}(s)) \vee \theta_{2} (\sigma) = \varepsilon ) \\
	& \wedge ... \wedge \theta _n (s) \in L(\cS_n/\cP^n) \wedge \theta _n (s) \theta _n (\sigma) \in L(\cP^n) \\
	& \wedge (\sigma \in \cS_n (\theta_{n}(s)) \vee \theta_{n} (\sigma) = \varepsilon ) \\
	\Leftrightarrow
	& \theta _1 (s\sigma) \in L(\cS_1/\cP^1) \wedge \theta _2 (s\sigma) \in L(\cS_2/\cP^2) \\
	& \wedge ... \wedge \theta _n (s\sigma) \in L(\cS_n/\cP^n) \\
	& (\mbox{by Equation (\ref{Equation6})}) .
\end{align*}
\hfill \bull

Note that 
\lfteqn
L(\wedge \cS_j/\cP) = & \theta _1 ^{-1} (L(\cS_1/\cP^1)) \cap \theta _2 ^{-1} (L(\cS_2/\cP^2)) \\
& \cap ... \cap \theta _n ^{-1} (L(\cS_n/\cP^n)).
\ndeqn
can also be written as
\lfteqn
L(\wedge \cS_j/\cP) = L(\cS_1/\cP^1) || L(\cS_2/\cP^2) || ... || L(\cS_n/\cP^n).
\ndeqn

Assume that $K \subseteq L(\cP)$ is conditionally decomposable \cite{komenda2012conditional} in the following sense:
$$
K= \theta _1 (K) || \theta _2 (K) || ... || \theta _n (K).
$$
Then, we have the following theorem.

\begin{theorem}
	There exist $n$ modular controllers $\cS _j$, $j=1,2,...,n$ such that 
	\lfteqn
	L(\wedge \cS_j/\cP)=K \\
	\wedge L(\cS_j/\cP^j) \subseteq \theta _j (K), j=1,2,...,n
	\ndeqn
	if and only if $\theta _j (K)$ is F-controllable with respect to $L(\cP^j)$, $\Sigma _c^j$, and $\Sigma _f^j$, for $j=1,2,...,n$.
	
\end{theorem}
{\em Proof}

We prove the ``if'' part and the ``only if'' part separately as follows. \\		
(IF) Suppose that $\theta _j (K)$ is F-controllable with respect to $L(\cP^j)$, $\Sigma _c^j$, and $\Sigma _f^j$. Then, there exist modular controllers $\cS _j$ such that $L(\cS_j/\cP^j)=\theta _j (K)$, for $j=1,2,...,n$. By Theorem 1,
\lfteqn
L(\wedge \cS_j/\cP) & = &  \theta _1 ^{-1} (L(\cS_1/\cP^1)) \cap \theta _2 ^{-1} (L(\cS_2/\cP^2)) \\
& & \cap ... \cap \theta _n ^{-1} (L(\cS_n/\cP^n)) \\
& = & L(\cS_1/\cP^1) || L(\cS_2/\cP^2) || ... || L(\cS_n/\cP^n) \\
& = & \theta _1 (K) || \theta _2 (K) || ... || \theta _n (K) \\
& = & K.
\ndeqn
Furthermore, 
\lfteqn
L(\cS_j/\cP^j)=\theta _j (K) \subseteq \theta _j (K), j=1,2,...,n
\ndeqn

(ONLY IF) Assume that there exist $n$ modular controllers $\cS _j$, $j=1,2,...,n$ such that 
\lfteqn
L(\wedge \cS_j/\cP)=K \\
\wedge L(\cS_j/\cP^j) \subseteq \theta _j (K), j=1,2,...,n .
\ndeqn
By Theorem 1, 
\lfteqn
& L(\wedge \cS_j/\cP)=K \\
\Rightarrow 
& K = L(\cS_1/\cP^1) || L(\cS_2/\cP^2) || ... || L(\cS_n/\cP^n)\\
\Rightarrow 
& \theta _j (K) = \theta _j (L(\cS_1/\cP^1) || L(\cS_2/\cP^2) || ... || L(\cS_n/\cP^n)) \\
\Rightarrow 
& \theta _j (K) \subseteq \theta _j (L(\cS_j/\cP^j)) = L(\cS_j/\cP^j) .
\ndeqn
Since $L(\cS_j/\cP^j) \subseteq \theta _j (K)$ by the assumption, we have
$$
L(\cS_j/\cP^j) = \theta _j (K), j=1,2,...,n .
$$
Therefore, $\theta _j (K)$ is F-controllable with respect to $L(\cP^j)$, $\Sigma _c^j$, and $\Sigma _f^j$, for $j=1,2,...,n$.

\hfill \bull

Let us assume that $K_j = \theta _j (K) \subseteq L(\cP^j)$ is  generated by a sub-automaton $\cH_j \sqsubseteq
\cP^j$, that is, $K_j = L(\cH_j)$ for some

\Lfteqn \label{Equation9A}
\cH_j =( Q^j_H, \Sigma ^j , \delta ^j_H, q_o^j ), \ j=1,2,...,n, \\
\Ndeqn

where $Q_H^j \subseteq Q^j$ and $\delta_H^j = \delta ^j |_{Q_H^j \times \Sigma_j}$.

We use the methods similar to those proposed in the previous section (with $\cH$ replaced by $\cH_j$) to obtain

\Lfteqn \label{Equation9B}
\cH^\uparrow_j=( Q_H^{\uparrow,j}, \Sigma , \delta_H^{\uparrow,j} , q_o^j ). \\
\Ndeqn

Consider the following modular controllers, for $j=1,2,...,n, s \in L(\cP^j)$,
\Lfteqn \label{Equation9}
\cS^{\uparrow \diamond}_j (s) & = \{\sigma \in \Sigma ^j :  s \sigma \in L(\cH^\uparrow_j) \} \\
& = \{\sigma \in \Sigma ^j :  \delta_H^{\uparrow,j} (q_o^j,s\sigma ) \in Q_H^{\uparrow,j} \} \\
\Ndeqn

\begin{figure*} [h!]
	\centering
	\includegraphics[keepaspectratio=true,angle=0, width= 0.7 \linewidth]{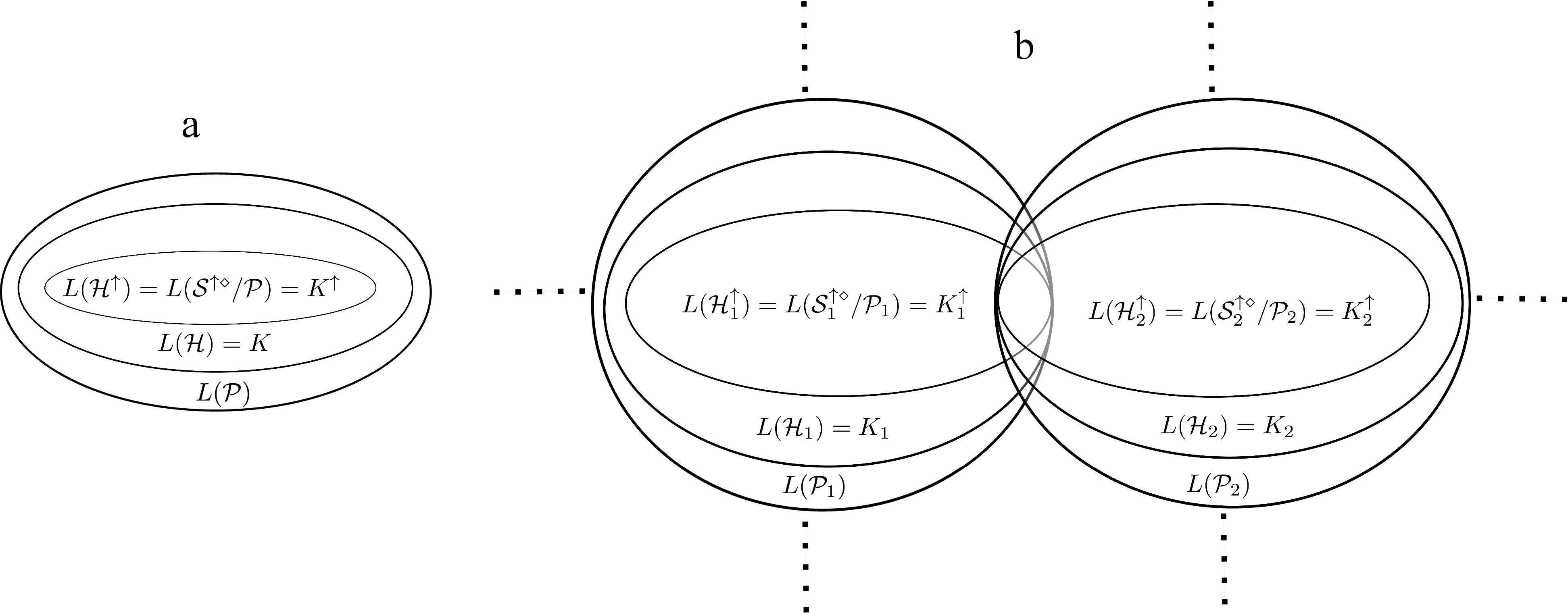}
	\caption{Subset diagrams of centralized control (a) and modular control (b).}
	\label{conceptual_of_two_languages}
\end{figure*}

\begin{table*}[h!]
	
	\caption{Events, labels, definitions, and attributes of automata in Fig. \ref{Basic_Model_Figure}.}
	\vspace*{-\abovedisplayskip}
	\label{Table 2}
	\footnotesize
	\begin{center}
		\begin{tabular}{|l l l l l l l |} 
			
			\hline
			Component &  event & label & definition & Controllable ($\Sigma_c$))& Forcible($\Sigma_f$) & forcing mechanism \\ 
			&  &  &  &  /Uncontrollable ($\Sigma_{uc}$)& /Unforcible ($\Sigma_{uf}$) &\\
			[0.5ex] 
			\hline\hline
			Load $j$ & $\mu^j_1$ &	$ej$ & Load $j$ is tripped & $\Sigma_{uc}$ & $\Sigma_{uf}$ & \\
			& $\mu^j_2$ &	$fj$ & Load $j$ is changed & $\Sigma_{c}$ & $\Sigma_{f}$ & Load $j$ shedding \\  
			& $\mu^j_3$ &	$gj$ & Load $j$ is back on line & $\Sigma_{uc}$ & $\Sigma_{uf}$ & \\ 
			\hline
			Line $k$ & $\beta^k_1$ & $kk$  & Line $k$ is tripped &  $\Sigma_{uc}$ & $\Sigma_{uf}$  & \\
			& $\beta^k_2$ &	$uk$  & Loading on Line $k$ is changed &  $\Sigma_{uc}$ & $\Sigma_{uf}$  &  \\
			& $\beta^k_3$ &	$hk$  & Line $k$ is back on line &  $\Sigma_{uc}$ & $\Sigma_{uf}$  & \\
			\hline
			Generator $i$ & $\gamma^i_1$ &	$ai$ & Generator $i$ is tripped &  $\Sigma_{uc}$ & $\Sigma_{uf}$ & \\
			& $\gamma^i_2$ &	$bi$ & Loading on Generator $i$ is changed &  $\Sigma_{c}$ & $\Sigma_{f}$ & Generator $i$ re-dispatch \\
			& $\gamma^i_3$ &	$ci$ & Generator $i$ is back on line &  $\Sigma_{uc}$ & $\Sigma_{uf}$ & \\
			\hline

		\end{tabular}
	\end{center}
\end{table*}

\begin{theorem}
	The modular controllers $\cS^{\uparrow \diamond}_j$, $j=1,2,...,n$, of Equation (\ref{Equation9}), ensure the safety of the controlled system, that is, 
	\lfteqn
	L(\cS^{\uparrow \diamond}_j/\cP^j) = L(\cH^\uparrow_j) \subseteq K_j, j=1,2,...,n\\
	\wedge L(\wedge \cS^{\uparrow \diamond}_j/\cP) \subseteq K .
	\ndeqn
	
\end{theorem}
{\em Proof}

Since there are two results to be proved, we prove them one by one. Let us first prove that, for all $s \in L(\cP_j)$, 
\lfteqn
s \in L(\cS^{\uparrow \diamond}_j/\cP^j) \Leftrightarrow s \in L(\cH^\uparrow_j)
\ndeqn
by induction on the length $|s|$ of $s$ as follows.

\noindent {\em Induction Base:} Since $\varepsilon \in L(\cS^{\uparrow \diamond}_j/\cP^j)$ and $\varepsilon \in L(\cH^\uparrow_j)$, for
$|s| = 0$, that is, $s=\varepsilon$, we have
\lfteqn
s \in L(\cS^{\uparrow \diamond}_j/\cP^j) \Leftrightarrow s \in L(\cH^\uparrow_j)
\ndeqn

\noindent {\em Induction Hypothesis:} Assume that for all $s \in \Sigma^*$, $|s| \leq m$,
\lfteqn
s \in L(\cS^{\uparrow \diamond}_j/\cP^j) \Leftrightarrow s \in L(\cH^\uparrow_j)
\ndeqn

\noindent {\em Induction Step:} We show that for all $s \in \Sigma^*$, $\sigma \in \Sigma$, $|s\sigma| = m+1$,
\lfteqn
s \sigma \in L(\cS^{\uparrow \diamond}_j/\cP^j) \Leftrightarrow s \sigma \in L(\cH^\uparrow_j)
\ndeqn

Indeed,
\begin{align*}
	& s \sigma \in L(\cS^{\uparrow \diamond}_j/\cP^j) \\
	\Leftrightarrow
	& s \in L(\cS^{\uparrow \diamond}_j/\cP^j) \wedge s \sigma \in L(\cP^j) \wedge \sigma \in \cS^{\uparrow \diamond}_j (s)  \\
	& (\mbox{by Equation (\ref{Equation6})}) \\
	\Leftrightarrow
	& s \in L(\cS^{\uparrow \diamond}_j/\cP^j) \wedge s \sigma \in L(\cP^j) \wedge s \sigma \in L(\cH^\uparrow_j)  \\
	& (\mbox{by Equation (\ref{Equation9})}) \\
	\Leftrightarrow
	& s \in L(\cH^\uparrow_j) \wedge s \sigma \in L(\cP^j) \wedge s \sigma \in L(\cH^\uparrow_j)  \\
	& (\mbox{by Induction Hypothesis}) \\
	\Leftrightarrow
	& s \sigma \in L(\cH^\uparrow_j) .
\end{align*}

Since $L(\cH^\uparrow_j) = K^\uparrow_j \subseteq K_j$, we have
\lfteqn
L(\cS^{\uparrow \diamond}_j/\cP^j) = L(\cH^\uparrow_j) \subseteq K_j, j=1,2,...,n.
\ndeqn

Furthermore, by Theorem 1, 
\lfteqn
L(\wedge \cS^{\uparrow \diamond}_j/\cP) & = L(\cS^{\uparrow \diamond}_1/\cP^1) || L(\cS^{\uparrow \diamond}_2/\cP^2) || ... || L(\cS^{\uparrow \diamond}_n/\cP^n) \\
& = L(\cH^\uparrow_1) || L(\cH^\uparrow_2) || ... || L(\cH^\uparrow_n) \\
& \subseteq K_1 || K_2 || ... || K_n \\
& = \theta _1 (K) || \theta _2 (K) || ... || \theta _n (K) \\
& = K .
\ndeqn

\hfill \bull

Modular controllers with a limited lookahead policy of $M$ steps, denoted by $\cS ^M_{CLL,j}$, can be used to implement $\cS^{\uparrow \diamond}_j$ similar to the implementation of $\cS^{\uparrow \diamond}$ by $\cS ^M_{CLL}$.


The relations among languages discussed earlier are shown in Fig. \ref{conceptual_of_two_languages} as a subset of diagrams for both centralized control and modular control.

\section{Implementation and Simulations}

We implement and simulate the modular controllers for power systems to mitigate cascading failures in this section. We first construct the sub-system model for each node based on the components connected to it and its neighboring nodes. These components are: generators, transmission lines, and loads. 

\begin{figure} [h!]
	\centering
	\includegraphics[keepaspectratio=true,angle=0,width=0.5\linewidth]{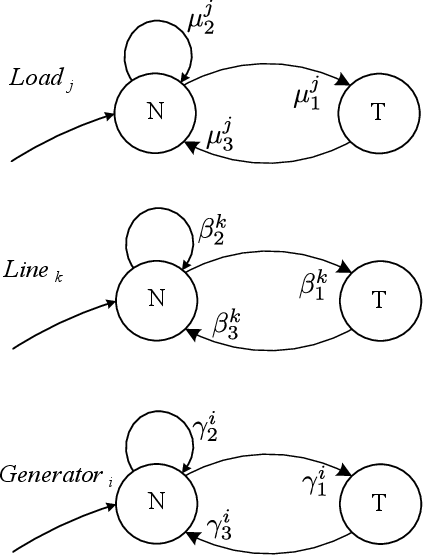}
	\caption{The automata for Load $i$, Line $k$, and Generator $j$. The states are normal (N) and tripped (T). The initial state (T) is denoted by $\rightarrow$. The events are defined in Table \ref{Table 2}. The subscript $j$ , $k$ and $i$ denote the enumerations of loads, lines, and generators respectively. }
	\label{Basic_Model_Figure}
\end{figure}

The automata for Load $i$, Line $k$, and Generator $j$ are shown in Fig. \ref{Basic_Model_Figure}. The events and their labels, definitions, and attributes are given in Table \ref{Table 2}.

%
%
%
%

\subsection{Sub-system model}

\begin{figure} [h!]
	\centering
	\includegraphics[keepaspectratio=true,angle=0, width= 0.8 \linewidth]{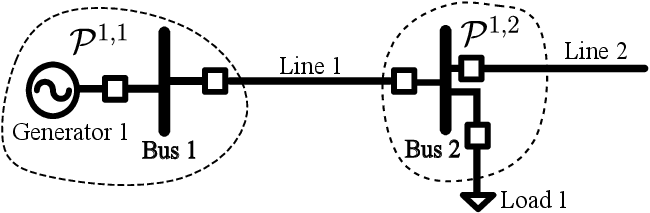}
	\caption{Simple power system with 2 nodes.}
	\label{2_Node_System}
\end{figure}

To design a modular controller $\cS_j$ for Node/Bus $j$ (we use ``bus'' and ``node'' interchangeably), we consider the neighboring nodes connected directly by a transmission line to Node $j$ and build the sub-system model $\cP^{j}$ for $\cS_j$. To illustrate how to do this, we consider a typical power system, partly shown in Fig. \ref{2_Node_System} with Node 1 and its neighboring node, Node 2. We use this example to illustrate the procedure developed in the previous section. The same procedure will be used in simulations to be discussed in Subsection D.

The automaton $\cP ^{1,1}$ for Node 1 is obtained by parallel composition of components connected to Node 1, namely, Generator 1 and Line 1, using the libFAUDES software \cite{libFAUDES}. To use libFAUDES, we label the events as shown in Table \ref{Table 2}. 

\begin{figure} [h!]
	\centering
	\includegraphics[keepaspectratio=true,angle=0, width= 0.95 \linewidth]{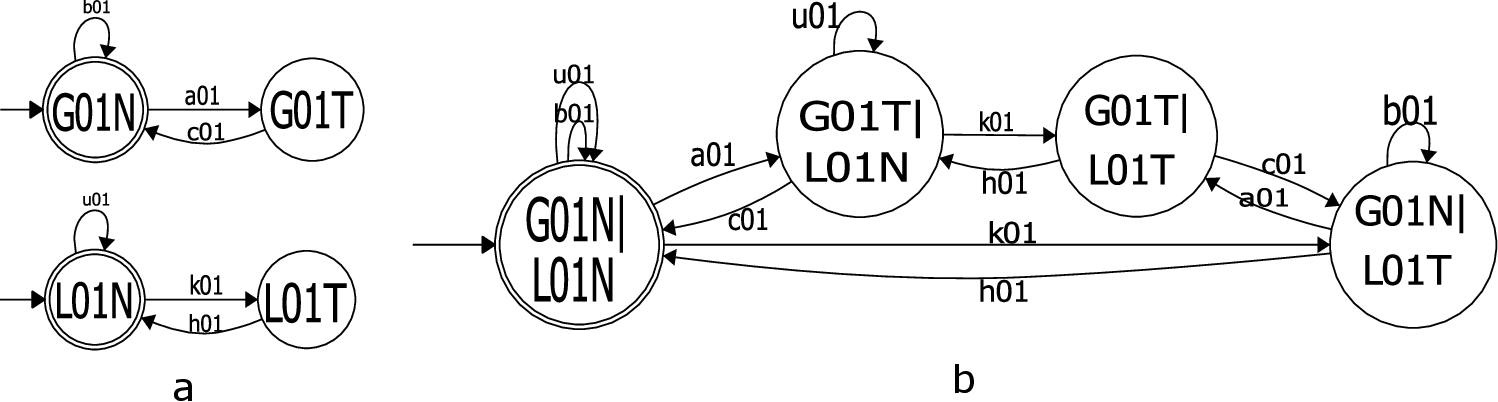}
	\caption{ The automaton $\cP ^{1,1}$ for Node 1 . a- Automaton models of individual components connected to Node 1. b- Parallel composition of all the components in (a).}
	\label{Node_1_Plant}
\end{figure}

The component automata, Generator 1 and Line 1, are shown in Fig. \ref{Node_1_Plant}-a. Generator 1 is denoted by G01, G01N means the normal state of Generator 1, and G01T means the tripped state of Generator 1. 
Similarly, Line 1 is denoted by L01, L01N means the normal state of Line 1, and N01T means the tripped state of Line 1.
The parallel composition $\cP ^{1,1}$ is shown in Fig. \ref{Node_1_Plant}-b. In the figure, a state of the parallel composition is labeled based on the corresponding states of its components.

\begin{figure*} [h!]
	\centering
	\includegraphics[keepaspectratio=true,angle=0, width= 0.95 \linewidth]{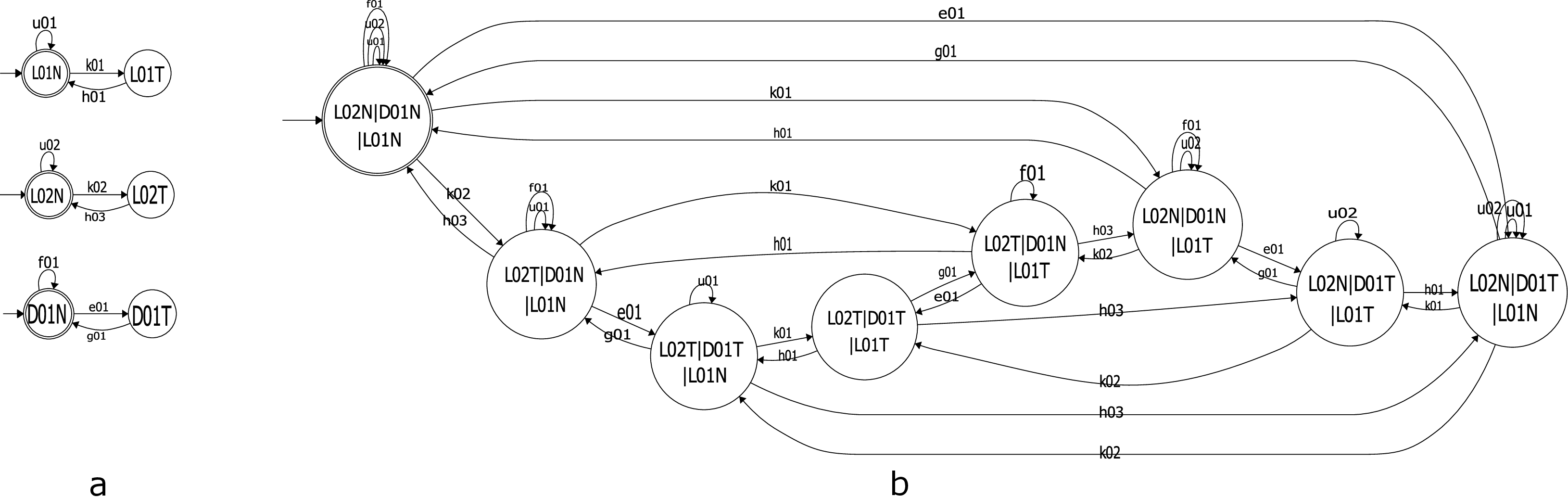}
	\caption{The automaton $\cP ^{1,2}$ for Node 2. a- Automaton models of individual components connected to Node 2. b- Parallel composition of all the components in (a).}
	\label{Node_2_Plant}
\end{figure*}

Using the same method, the automaton $\cP ^{1,2}$ for Node 2 shown in Fig. \ref{Node_2_Plant}-b is obtained by parallel composition of components connected to Node 2,  that is, Load 1, Line 1, and Line 2 shown Fig. \ref{Node_2_Plant}-a. 
In the figure, the states for load 1 are denoted by D01N and D01T for the normal and trip states, respectively.

\begin{figure*} [h!]
	\centering
	\includegraphics[keepaspectratio=true,angle=0, width= 0.95 \linewidth]{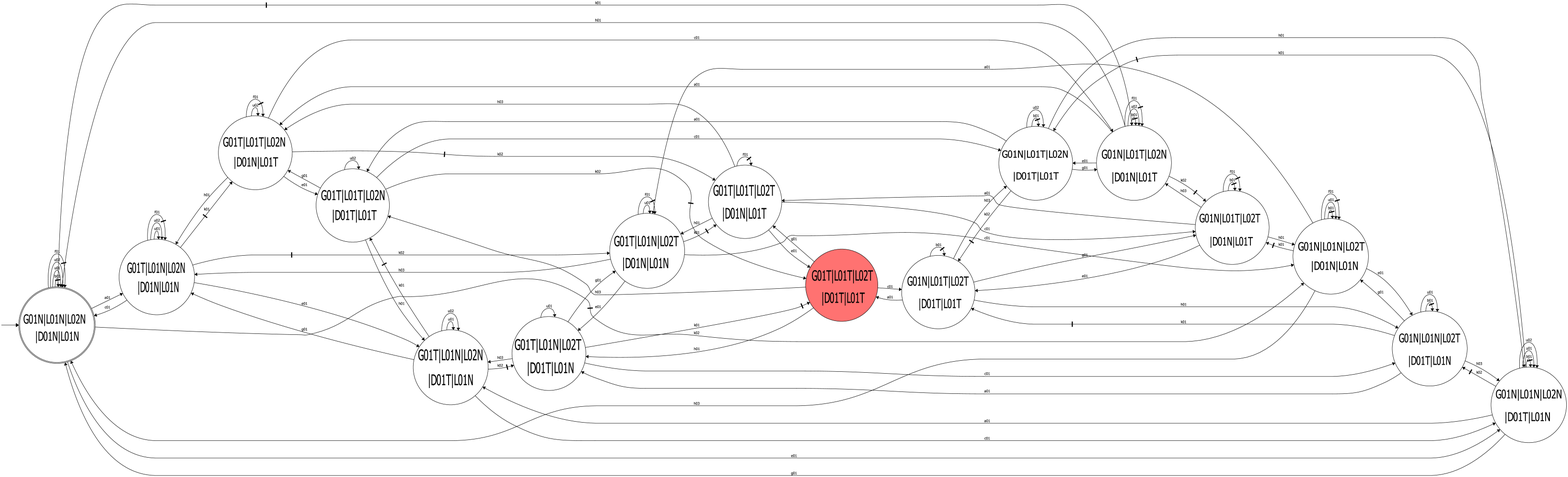}
	\caption{The sub-system model $\cP^{1} = \cP ^{1,1} || \cP ^{1,2}$. }
	\label{Node_1_2_Plant}
\end{figure*}

The sub-system model $\cP^{1}$ consists of Node 1 and its neighboring node, Node 2. Hence, $\cP^{1}$ is the parallel composition of $\cP ^{1,1}$ and $\cP ^{1,2}$:
$$
\cP^{1} = \cP ^{1,1} || \cP ^{1,2} =( Q^1, \Sigma ^1 , \delta ^1, q_o^1 ),
$$
which is shown in Fig. \ref{Node_1_2_Plant}. Note that this $\cP^{1}$ corresponds to the $\cP^{1}$ is Equation (\ref{Equation5}).

\subsection{Specification} 

A cascading failure can be defined as a string of uncontrollable events that lead the system from a legal/safe state to some illegal/unsafe states. In this example, the illegal state is
\lfteqn
D01T,L03T,G01T,L01T,L02T
\ndeqn
which are marked in red in Fig. \ref{Node_1_2_Plant}.

\begin{figure*} [h!]
	\centering
	\includegraphics[keepaspectratio=true,angle=0, width= 0.98 \linewidth]{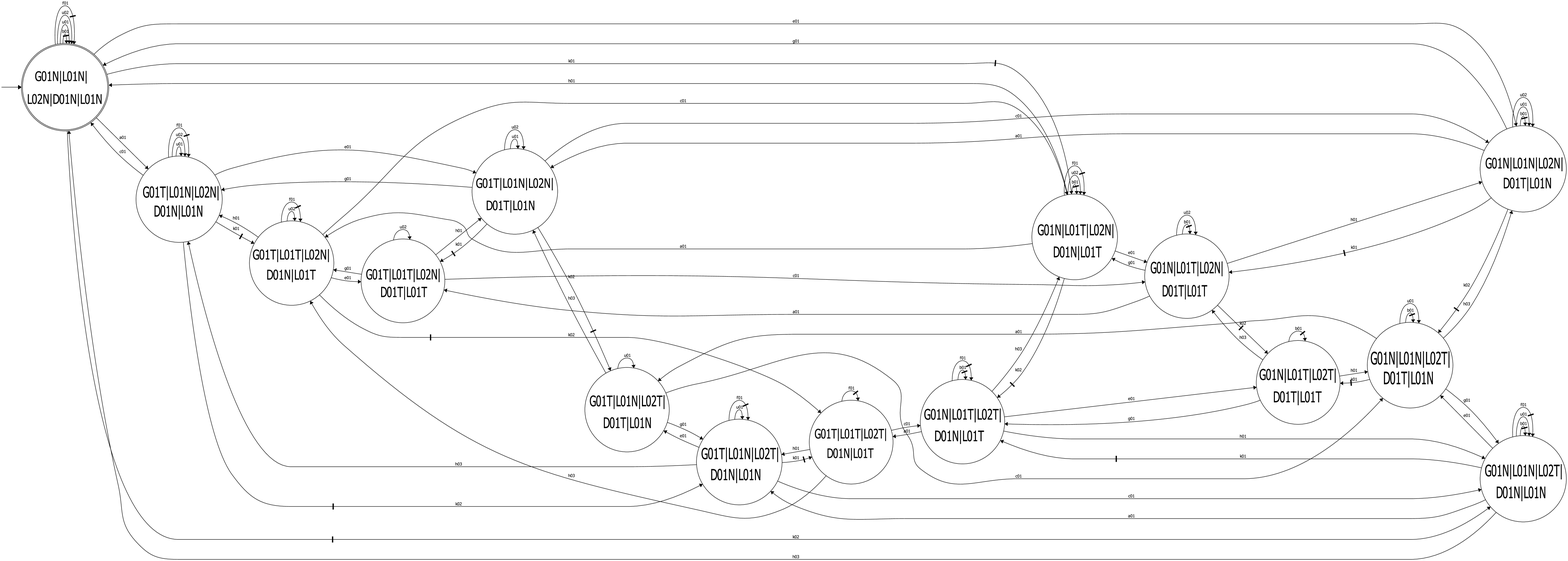}
	\caption{Specification automaton $\cH_{1}$ for $\cP^{1}$.}
	\label{specification_A1_A2}
\end{figure*}

Hence, the sub-automaton $\cH_1$ generating the specification language $K_1$ is obtained by removing the illegal states from $\cP_1$.
$$
\cH_1 =( Q^1_H, \Sigma ^1 , \delta ^1_H, q_o^1 ) 
$$
is shown in Fig. \ref{specification_A1_A2}. Note that this $\cH_1$ corresponds to the $\cH_1$ is Equation (\ref{Equation9A}).

Suppose there are $n$ buses in the power system. Then the entire power system is modeled by 
$$
\cP = \cP^{1}||\cP^{2} || ... || \cP^{n}.
$$
The global specification is given by 
$$
K= K_1 || K_2 || ... || K_n .
$$
It is known in DES \cite{komenda2007control, komenda2012conditional} that whenever $K= K_1 || K_2 || ... || K_n$, the conditional decomposability condition of $K$ is automatically satisfied. Therefore, we can use modular control to mitigate cascading failures.

%
%

\subsection{Controller synthesis}


We use libFAUDES software to synthesize controller $\cS_j$. To do so, the plant model $\cP_{j}$ and the specification language $K_j$ are needed, which are obtained in the previous two subsections. Additionally, controllability attributes need to be defined for events in the system. Table II shows the controllability attributes of the events (whether an event is controllable or forcible).

Note that transmission line tripping, which is event $kk$, are not controllable by themselves, but can be preempted by forcible events. Note also that load shedding, which is event $fj$, and generation re-dispatch, which is event $bi$, are both controllable and forcible.


The forcible actions adopted in this work can be implemented via the measures shown in Table \ref{Measures}, with the typical time scale for each action indicated accordingly. Note that in Table \ref{Measures}, although under frequency load shedding (UFLS) and under voltage load shedding  (UVLS) can be used as means of preventing cascading failures depending on voltage and frequency signals, they may not be used directly by the proposed approach, but rather, the proposed approach can be more fitted within special protection scheme (SPS). Another possible activation method via remedial action scheme (RAS) is automatic generation control (AGC). RAS or System integrity protection schemes (SIPS) may be used in distributed fashion to enhance resiliency and cybersecurity \cite{RavikumarSIPS}.

\begin{table*}[h!]
	
	\caption{Forcible Events (Measures) for Mitigating Cascading Failures and Corresponding mapped events}
	\vspace*{-\abovedisplayskip}
	\label{Measures}
	\footnotesize
	\begin{center}
		\begin{tabular}{|l l l l l|} 
			
			\hline
			Measure & Typical & Telemetry / & Mapped event & Remarks\\ 
			 & Timescale &  Control Path & & \\ 
			[0.5ex] 
			\hline\hline
			UFLS & 20 ms – few s &	Local relays & Not applicable & \\
			\hline
			UVLS & 0.5 – 10 s & Local voltage relays  & Not applicable & \\
			\hline
			SPS / RAS & 50 ms – 1 s &	Local + wide-area signals & $\{ fj, bi \}$ & Similar implemnation can be found in \cite{Madani5565539} and \cite{toro2023toward}\\
			\hline
			AGC & Seconds–minutes &	SCADA → EMS → SCADA & $bi$ & \\
			\hline
			Governor response & 0 – 10 s &	Local generator control & Not applicable & \\
			\hline
			FACTS / STATCOM & 20 ms – 300 ms &	Local high-speed control & Not applicable & \\
			\hline
			Controlled islanding & 100 ms – seconds &	PMUs + wide-area comms & Not applicable & \\
			\hline
			Operator actions & Minutes & SCADA / EMS & Not applicable & \\
			\hline
		
		\end{tabular}
	\end{center}
\end{table*}

\par After knowing the controllability attributes and the specification language, a controller can be synthesized for Node 1 by applying the method outlined in Section III. Since events leading to the illegal state can be preempted by forcible events,  $K_1 = L(\cH_1)$ is F-controllable. Hence, the sub-automaton in Equation (\ref{Equation9B}) is $\cH^\uparrow_1 = \cH_1$. By Equation (\ref{Equation9}), the controller is given by, for $w \in \theta_1
(L(\cP^1))$,

\lfteqn
\cS^{\uparrow \diamond} _1 (w) & = \{\sigma \in \Sigma ^1 :  \delta_H^{\uparrow,1} (q_o^j,s\sigma ) \in Q_H^{\uparrow,1} \} \\
& = \{\sigma \in \Sigma ^1 : \delta ^1_H (q_o^1 , w\sigma) \in Q^1_H \} \\
& = \cS^{\diamond} _1 (w) .
\ndeqn

\begin{figure*} [h!]
	\centering
	\includegraphics[keepaspectratio=true,angle=0, width= 0.95 \linewidth]{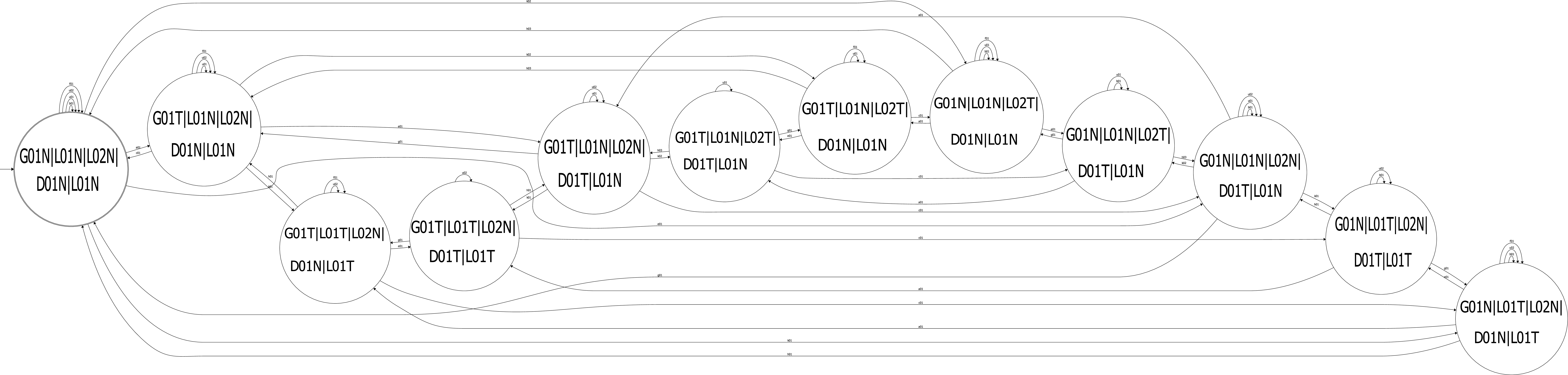}
	\vspace{-10pt}
	\caption{$\cS_{1}$: The realization of $\cS^{\diamond} _1$ of Node 1.}
	\label{Supervisor_A1_A2}
\end{figure*}

\begin{figure} [h!]
	\centering
	\includegraphics[keepaspectratio=true,angle=0, width= 0.4 \linewidth]{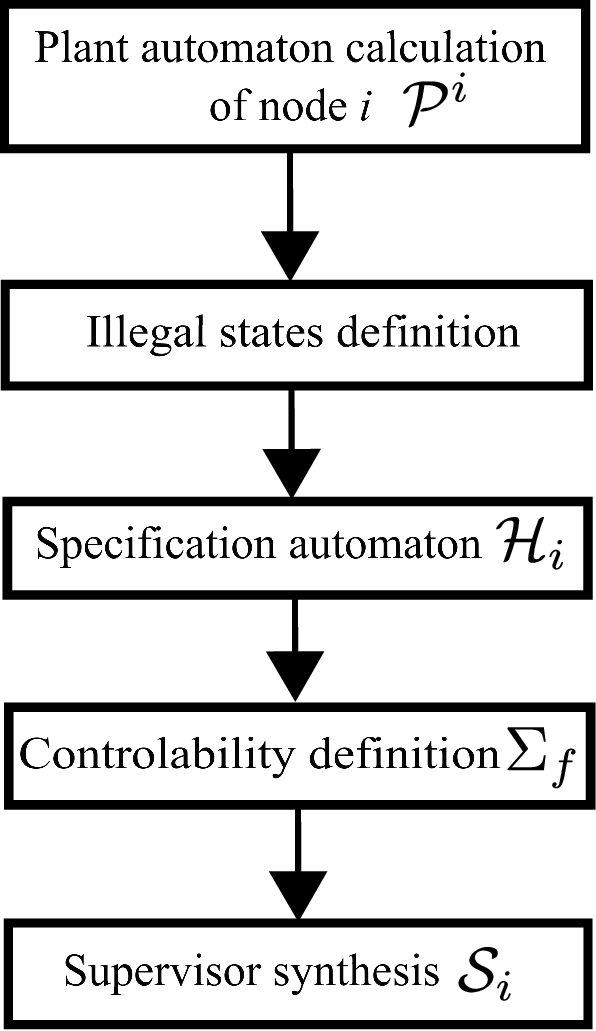}
	\vspace{-10pt}
	\caption{Flowchart of the supervisor synthesis procedure.}
	\label{flowchart_1}
\end{figure}

The realization of $\cS^{\diamond} _1$ as an automaton 
\lfteqn
R^{1} = (Y^{1}, \Sigma, g, y_{0}^{1}) 
\ndeqn
can be synthesized based on the sub-system model $\cP^{1}$ in Fig. \ref{Node_1_2_Plant} and the specification automaton $\cH_{1}$ in Fig. \ref{specification_A1_A2} using libFAUDES software. The synthesis algorithm is based on \cite{cassandras2009introduction}. Since libFAUDES is developed for conventional supervisory control without forcible events, we need to modify event attributes as shown in Table III to use libFAUDES. 
For example, we view $ej$ as controllable, because it can be preempted by $fj$. The resulting $R^{1}$ is shown in Fig. \ref{Supervisor_A1_A2}.
The supervisor synthesis function requires two inputs: The plant automaton $\cP^{1}$ with the modified controllability attributes shown in Table III and the specification automaton $\cH_{1}$. Its output is the supervisor automaton $\cS_{1}$ or $R^1$. A flow chart of the procedure described above is shown in Fig. \ref{flowchart_1}.

\begin{table}[h!]
	
	\caption{Modified events Attributes}
	\vspace*{-\abovedisplayskip}
	\label{Table 3}
	\footnotesize
	\begin{center}
		\begin{tabular}{|l c c c |} 
			
			\hline
			Component &  event & Controllable ($\Sigma_c$))& Forcible($\Sigma_f$) \\ 
			&  &  /Uncontrollable ($\Sigma_{uc}$)& /Unforcible ($\Sigma_{uf}$)\\
			[0.5ex] 
			\hline\hline
			Load $j$ &	$ej$ & $\Sigma_{c}$ & $\Sigma_{uf}$ \\
			&	$fj$ & $\Sigma_{c}$ & $\Sigma_{f}$ \\  
			&	$gj$ & $\Sigma_{uc}$ & $\Sigma_{uf}$ \\ 
			\hline
			Line $k$ &	$kk$  &  $\Sigma_{c}$ & $\Sigma_{uf}$  \\
			&	$uk$  &  $\Sigma_{uc}$ & $\Sigma_{uf}$  \\
			&	$hk$  &  $\Sigma_{uc}$ & $\Sigma_{uf}$  \\
			\hline
			Generator $i$ &	$ai$ &  $\Sigma_{c}$ & $\Sigma_{uf}$ \\
			&	$bi$ &  $\Sigma_{c}$ & $\Sigma_{f}$ \\
			&	$ci$ &  $\Sigma_{uc}$ & $\Sigma_{uf}$ \\
			\hline

		\end{tabular}
	\end{center}
\end{table}

\subsection{Simulation results}

\begin{figure} [h!]
	\centering
	\includegraphics[keepaspectratio=true,angle=0, width= 0.7 \linewidth]{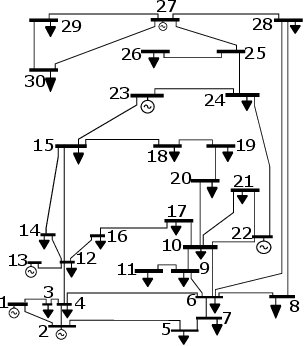}
	\vspace{-10pt}
	\caption{Single-Line diagram of the IEEE 30-bus system.}
	\label{30_bus}
\end{figure}

\begin{figure*} [h!]
	\centering
	\includegraphics[keepaspectratio=true,angle=0, width= 0.8 \linewidth]{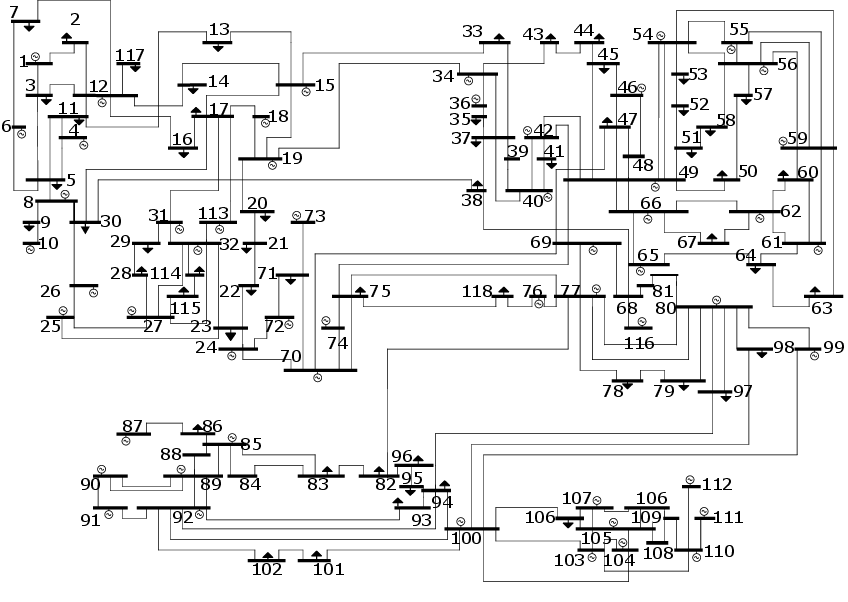}
	\vspace{-10pt}
	\caption{Single-Line diagram of the IEEE 118-bus system.}
	\label{118_bus}
\end{figure*}

\begin{figure*} [h!]
	\centering
	\includegraphics[keepaspectratio=true,angle=0, width= 0.8 \linewidth]{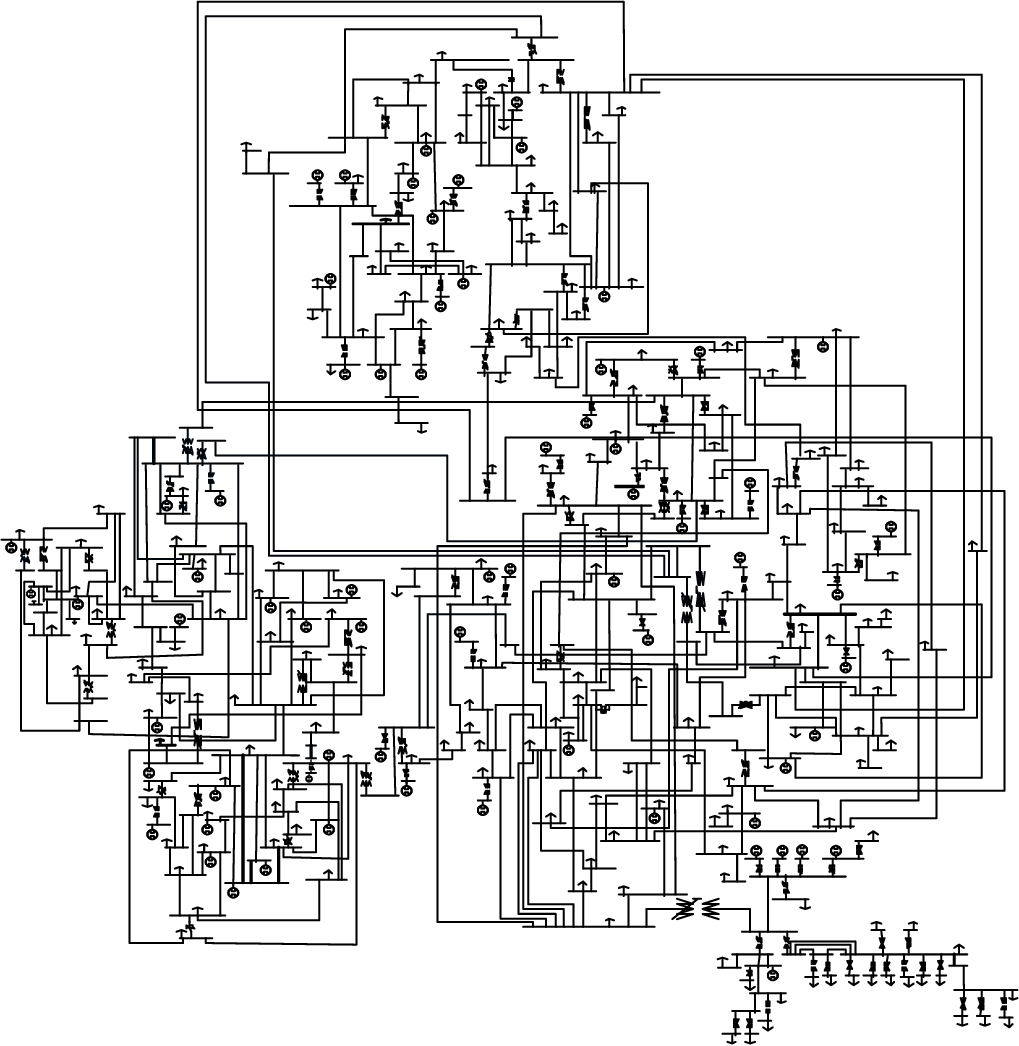}
	\vspace{-1em}
	\caption{Single-Line diagram of the IEEE 300-bus system \cite{Matpower} \cite{IEEE300}.}
	\label{300_bus_single_line}
\end{figure*}

The procedure followed for the illustrative example in subsections V-A, V-B, and V-C is implemented and simulated for the IEEE 30-bus, 118-bus, and 300-bus systems \cite{Matpower} \cite{IEEE300} to verify the findings of the proposed modular control. The single-line diagrams of the IEEE 30-bus, 118-bus, and IEEE 300-bus systems are shown in Figs. \ref{30_bus},  \ref{118_bus}, and  \ref{300_bus_single_line}, respectively.
It is assumed in this paper that nodes send and receive information to direct neighbors without any time delay. The simulation time, is, however, considered and recorded, which includes the computation time of the optimal control actions and the supervisors' iterators.

The simulation setup is done in MATLAB environment, which is used to simulate the system with Matpower \cite{Matpower} and DCSIMSEP \cite{EppesteinRandomChemisrty, dcsimsep}; and to send and receive information from the controllers. 
The model used in the paper is comparable to the OPA model \cite{Complexdynamics}, specifically the fast phase process part, where DC power flow is used and generation redsipatch with linear programming is implemented. It is worth noting that real power flows (with real power load shedding and real power generation redispatch) are used to verify the proposed modular control method, in which DC approximation is a reasonable approach. Nevertheless, the developed DES based modular supervisory control framework is not limited to DC modeling approach. In future work, we will extend our work to AC power studies and other forcible measures such as control of reactive power compensation devices.

The controllers are implemented in MATLAB, where the DES operations functions are imported from C++ libFAUDES \cite{libFAUDES} DES library into MATLAB, and as illustrated in Subsections A, B and C.
The controllers are synthesized based on the specification and plant models. 
As mentioned earlier, each node or bus in the power system has its controller; for example, the IEEE 300-bus system has 300 modular controllers. To optimally control the amount of load shedding and generation re-dispatch, each of the modular controllers solves the following optimization problem at a given node $i$ if the DES controller for that bus allows it.
\Lfteqn 
min \underset{x^{i}} \: \sum_{j} x_j^{i}
\\
s.t.
\\
\sqrt{y_t^{i}}((A^{T}_{t})^{-1})x^{i} \leqslant F_{Max}^{i} - (P)^{i}
\\	
D^{i} x^{i} \leqslant (F(l_c)_{Max})^{i} - P(l_c)^{i},
\Ndeqn
where $x_j$ are the amounts of load to be shed, $y_t$ is the diagonal matrix of the line admittance, $A_{t} = \sqrt{y_t^{i}A^{i}}$ and $A$ is the line-node incidence matrix, $F_{Max}$ are the lines max capacities, 
$P$ are the line power flows, $D = (PTDF_{i,lc})$ is the value of the power transfer distribution factor ($PTDF$) of the buses for the critical transmission line $l_c$, and $l_c$ is the most loaded line in site $i$. The superscript $i$ refers to bus $i$ and connected components to its neighborhood, i.e., direct components such as lines, loads, and generators that are connected to bus $i$ and lines, loads, and generators that are connected to neighboring buses to bus $i$, as illustrated in the previous subsection. PTDF is defined as the sensitivity matrix of the power flow of lines to real power injection at the buses of the power system. The optimization problem for each site $i$ is solved for each site once the higher level DES supervisor allows it.

\begin{figure} [h!]
	\centering
	\includegraphics[keepaspectratio=true,angle=0, width= 0.6 \linewidth]{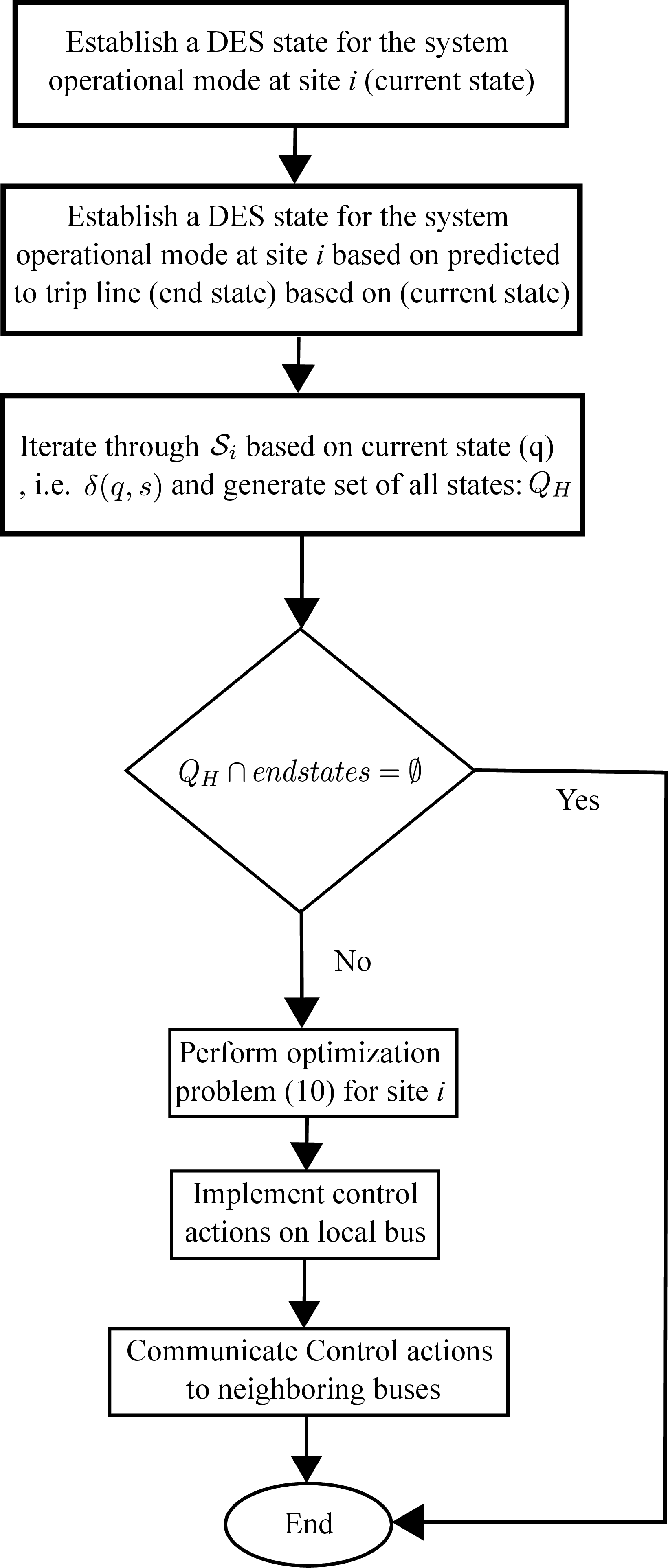}
	\vspace{-10pt}
	\caption{Flowchart of the implementation of the modular DES controller.}
	\label{flowchart_2}
\end{figure}

A flow chart of the procedure described above is shown in Fig. \ref{flowchart_2}. The predicted trip line is determined as the most loaded line in the neighborhood of site $i$. The algorithm in Fig. \ref{flowchart_2} is triggered whenever a line trip happens in the system.

The framework developed in this paper has been applied to the three case studies in two steps as previously shown. The first step is summarized in Fig. \ref{flowchart_1}, while the second step is illustrated in Fig. \ref{flowchart_2}.

\begin{figure} [h!]
	\centering
	\includegraphics[keepaspectratio=true,angle=0, width= 1 \linewidth]{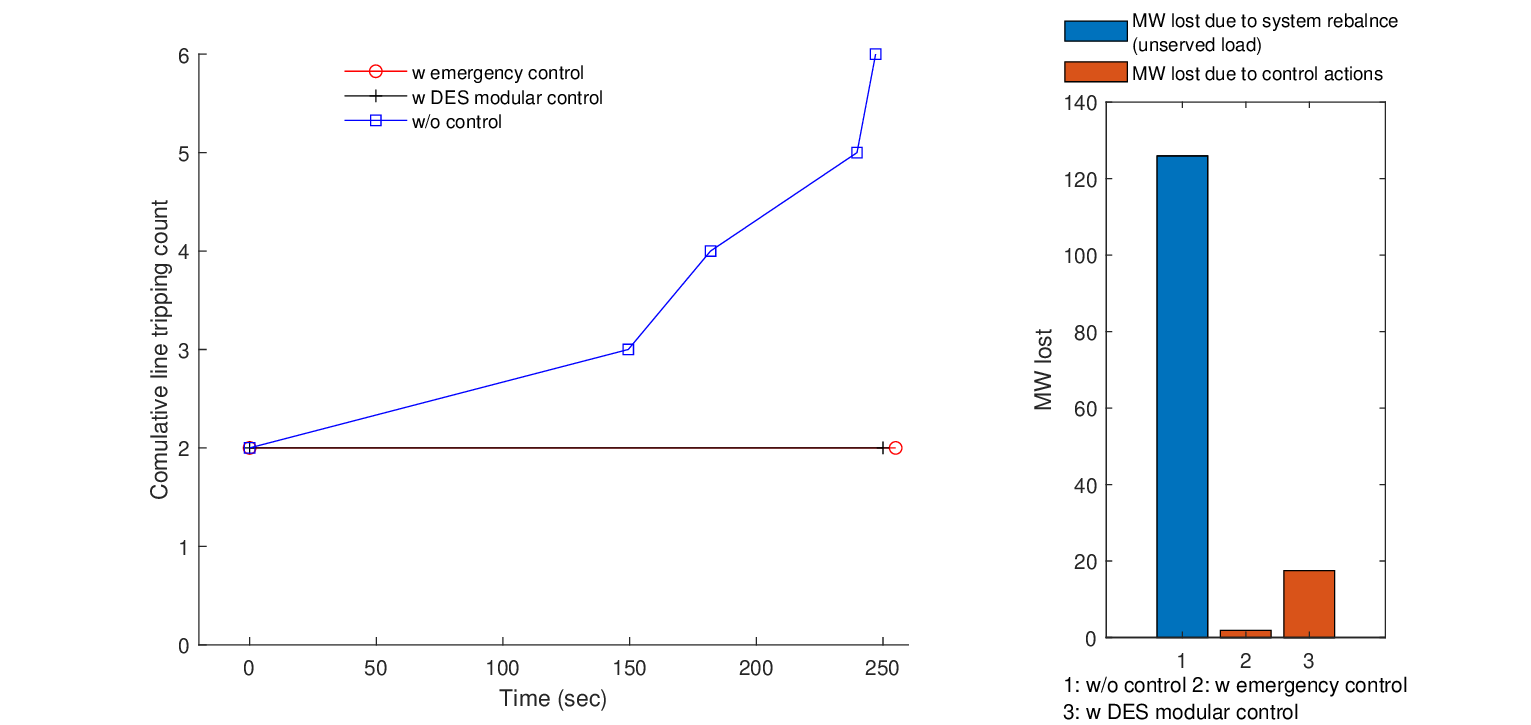}
	\vspace{-20pt}
	\caption{Effect of applying the modular DES control approach for the IEEE 30-bus system.}
	\label{30_bus_single_case}
\end{figure}

\begin{figure} [h!]
	\centering
	\includegraphics[keepaspectratio=true,angle=0, width= 1 \linewidth]{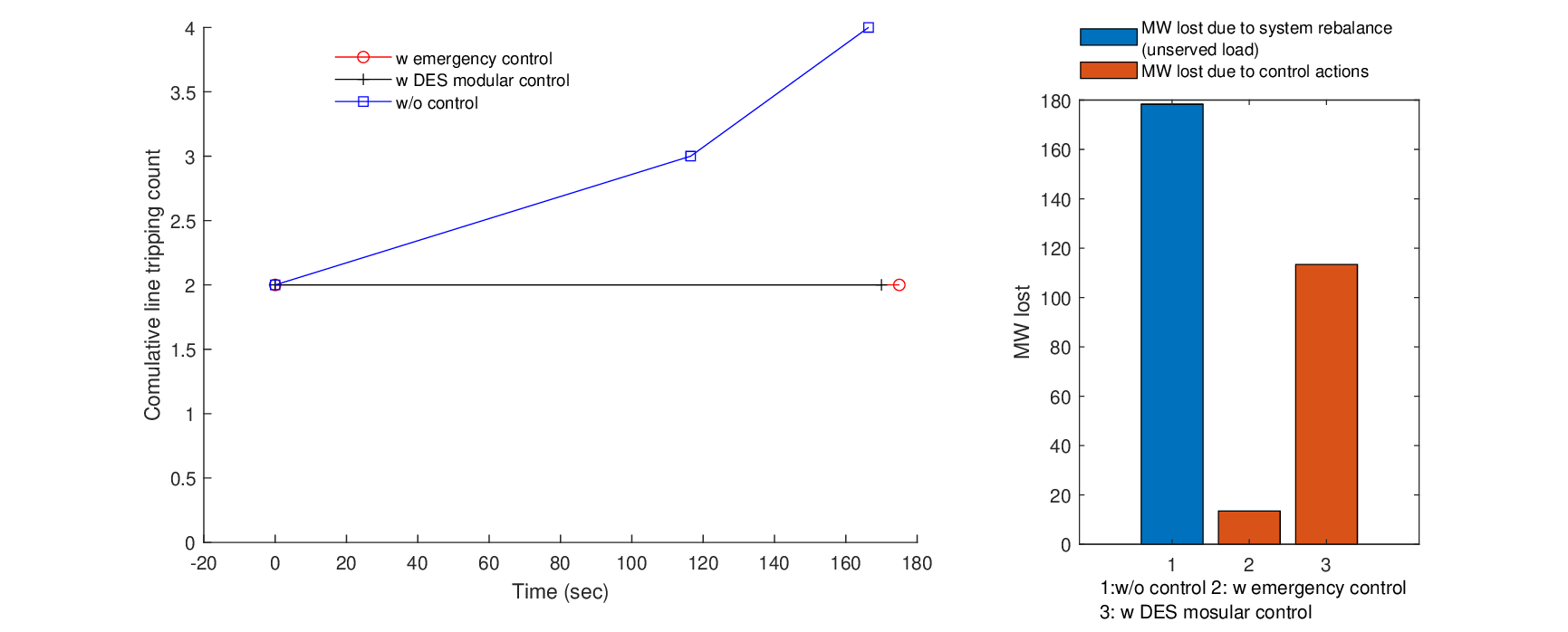}
	\vspace{-20pt}
	\caption{Effect of applying the modular DES control approach for the IEEE 118-bus system.}
	\label{118_bus_single_case}
\end{figure}

\begin{figure} [h!]
	\centering
	\includegraphics[keepaspectratio=true,angle=0, width= 1 \linewidth]{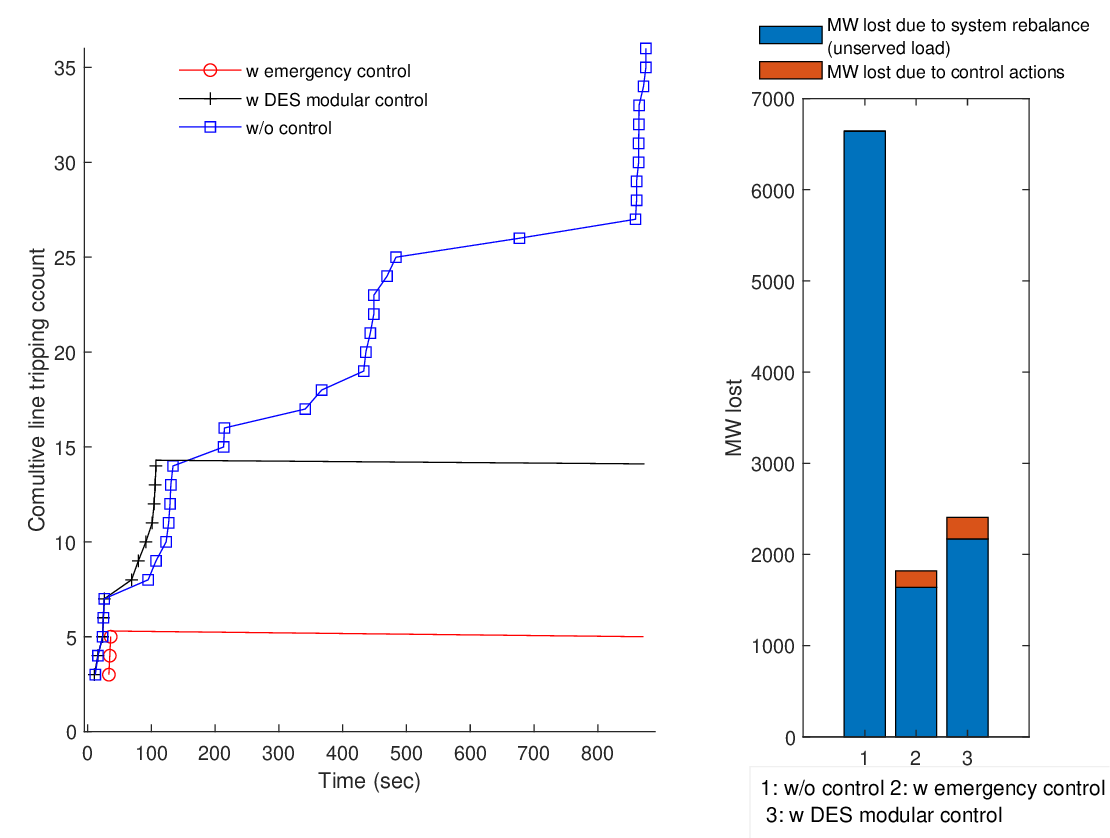}
	\vspace{-2em}
	\caption{Effect of applying the modular DES control approach after lines pair $ \{(23,39)\}$ initial trip.}
	\label{300_Bus_specific_case}
\end{figure}

Fig. \ref{30_bus_single_case} shows the results of specific scenarios from the IEEE 30-bus system simulation results. The $N-2$ contingency simulated was the line pair $ \{(34,37)\}$. The modular control approach stopped the failure cascading, but with higher MW lost than the LLP DES control method discussed in \cite{LLPWasseem}. The MW lost index in the bar chart in Fig. \ref{30_bus_single_case} is divided into two components: the MW lost due to unserved load because of successive line trips, during this process, the power system may perform rebalance function between load and generation, and the MW lost due to control actions taken by the controller trying to prevent cascading failure. Control actions include load shedding and generation redispatch that may result in another rebalance operation. It is observed in the figure that in the case of no control, no MW lost due to control actions are observed, and only the MW lost due to rebalance. While in the emergency control and the adopted DES method, we often only observe the MW lost due to control actions, while the MW lost due to rebalance can happen as well, as shown in Fig. \ref{300_Bus_specific_case}. The higher MW lost is because, in modular control, the modular supervisors can only make decisions based on the information received from neighboring nodes and send control actions to the neighboring nodes, making the solution local and not global.
To further verify our approach, we performed simulation studies on the IEEE-118 bus system. A specific case of the simulation studies is shown in Fig. \ref{118_bus_single_case}, line pair $ \{(142,143)\}$ were tripped as the initial $N-2$ trip.
A typical simulation of the IEEE 300-bus system is shown in Fig. \ref{300_Bus_specific_case}, where the $N-2$ contingency of the line pair $ \{(23,39)\}$ is simulated.  
The proposed modular control stops the failure cascading as desired, but with a higher MW loss than the emergency control method discussed in \cite{thesis}, which is a centralized control method that uses a global optimization for the power system. 
The higher MW lost is because, in the modular control, the modular controllers can only make decisions based on the information received from neighboring nodes and send control actions to the neighboring nodes, making the solution local and not global. Nevertheless, as discussed in Section IV, the modular controller addresses the issues that a centralized controller has. Fig. \ref{300_Bus_specific_case} shows the simulation results of both the emergency control and the DES modular control methods. The MW lost due to the system rebalance can be linked to the line tripping counts on the left side in the figure for the same case study. It is noticed in this case that the initial control actions after the initial $N-2$ contingency did not immediately stop the failure cascade, and further control actions were taken after each additional line trip occurred. This resulted in additional MW lost due to rebalance.

Monte Carlo simulations have been carried out for the three adopted systems to further verify the proposed approach. In the simulations, the $N-2$ initial tripping follows a uniform random distribution. 

\begin{figure} [h!]
	\centering
	\includegraphics[keepaspectratio=true,angle=0, width= 0.8 \linewidth]{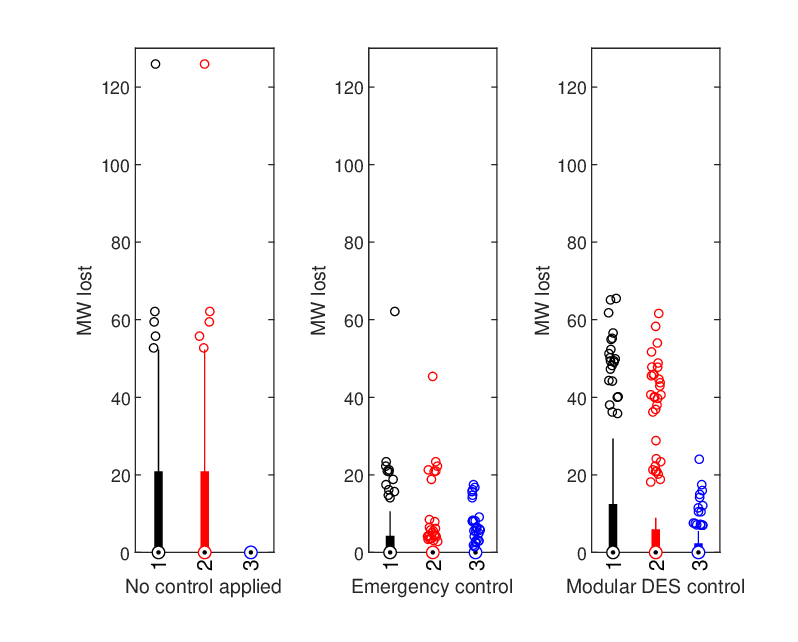}
	\vspace{-10pt}
	\caption{Monte Carlo simulation applying the modular DES control approach for the IEEE 30-bus system. 1-Total MW lost (the sum of both for each simulated case), 2-MW lost due to rebalance, and 3-MW lost due to control actions.}
	\label{30_Bus_MW_Lost}
\end{figure}

\begin{figure} [h!]
	\centering
	\includegraphics[keepaspectratio=true,angle=0, width= 0.8 \linewidth]{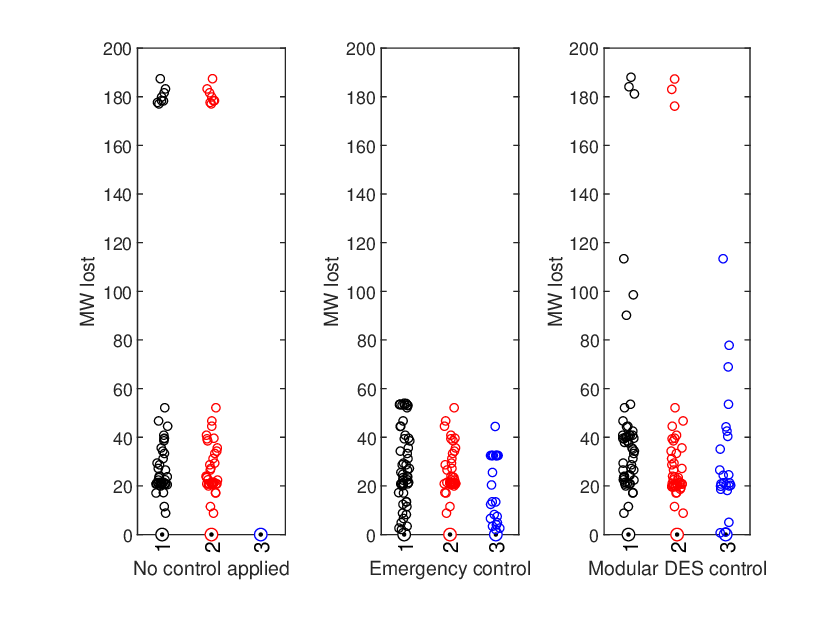}
	\vspace{-10pt}
	\caption{Monte Carlo simulation applying the modular DES control approach for the IEEE 118-bus system. 1-Total MW lost (the sum of both for each simulated case), 2-MW lost due to rebalance, and 3-MW lost due to control actions.}
	\label{118_Bus_MW_Lost}
\end{figure}

\begin{figure} [h!]
	\centering
	\includegraphics[keepaspectratio=true,angle=0, width= 0.8 \linewidth]{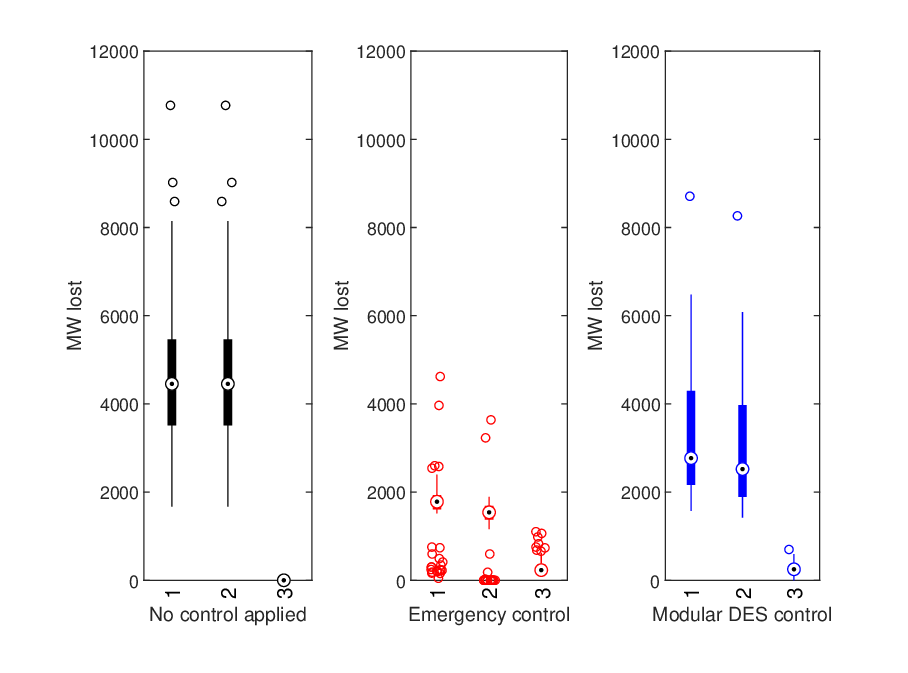}
	\vspace{-2em}
	\caption{Monte Carlo simulation applying the modular DES control approach for the IEEE 300-bus system. 1-Total MW lost (the sum of both for each simulated case), 2-MW lost due to rebalance, and 3-MW lost due to control actions.}
	\label{300_Bus_MW_Lost}
\end{figure}

The loads follow a normal distribution, with a standard deviation of 0.15 per unit from the nominal loads. Results of the simulations and the comparison with the emergency control method and without any control to mitigate the failure cascade are shown in Figs. \ref{30_Bus_MW_Lost} - \ref{300_Bus_MW_Lost}. Similar to specific cases analysis, we separated the MW lost into two categories: MW lost due to rebalance, and MW lost due to control actions. We also added the total MW lost, which is the sum of both for each simulated case. 
\par
For the IEEE 300-bus system, the median MW loss for the DES modular control method is 2770 MW. For the emergency control method, it is 1782 MW, and without using any control, it is 4454 MW. Our method shows that it could mitigate the failure cascade in most cases and reduce the overall lost MW in the system. 

As shown in the previous results, the emergency control results in less MW lost than the proposed modular DES approach. This is due to the fact that emergency control is a centralized controller, which requires collecting data from all the nodes of the system and sending control actions to all the nodes. In theory, this approach may be superior. Still, in practice, it may have some drawbacks: (1) The communication delay may impact the controller's performance, since each node has to send and receive information with the central controller. This delay is significantly reduced in modular control, as each node only needs to communicate directly with its surrounding nodes. (2) The overall architecture is less reliable than the modular control approach, since in centralized control, there can be a single point of failure.

The proposed method has been compared with other methods that are also based on modular control \cite{Reciprocally}, \cite{SHI2015582}, \cite{WARNIER201715}, while the authors used different indices to express the effectiveness of their proposed approaches. The results, in general, show similar behavior. It is also worth mentioning that in this paper we did not focus on the study of the communication delay between controlling agents on the effectiveness of the proposed approach.

\begin{figure} [h!]
	\centering
	\includegraphics[keepaspectratio=true,angle=0, width= 1 \linewidth]{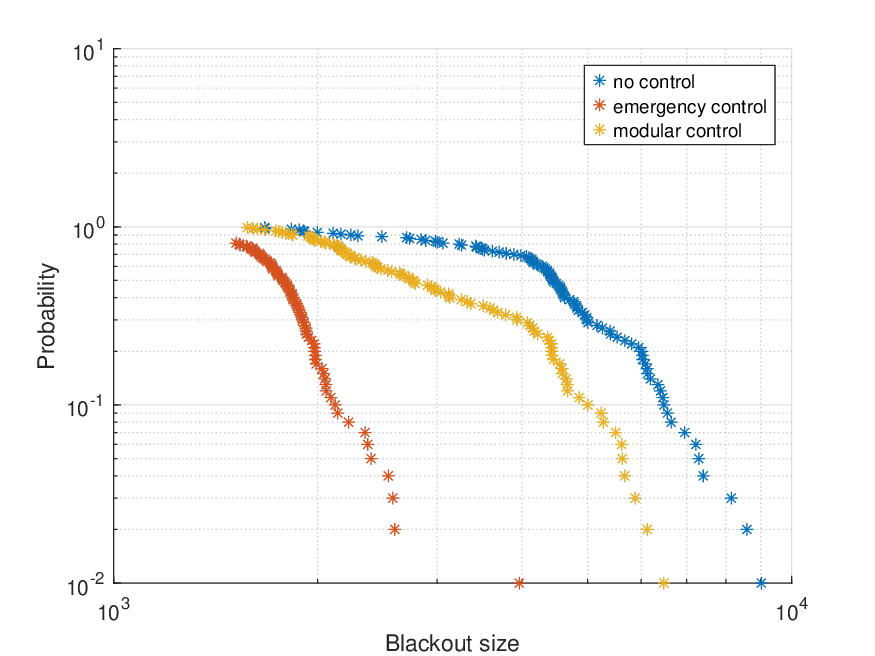}
	\vspace{-20pt}
	\caption{Complementary cumulative distribution (CCD) plot of the IEEE 300 system applying three scenarios for the blackout size measured in MW.}
	\label{loglog_plot_300}
\end{figure}

\begin{figure} [h!]
	\centering
	\includegraphics[keepaspectratio=true,angle=0, width= 1 \linewidth]{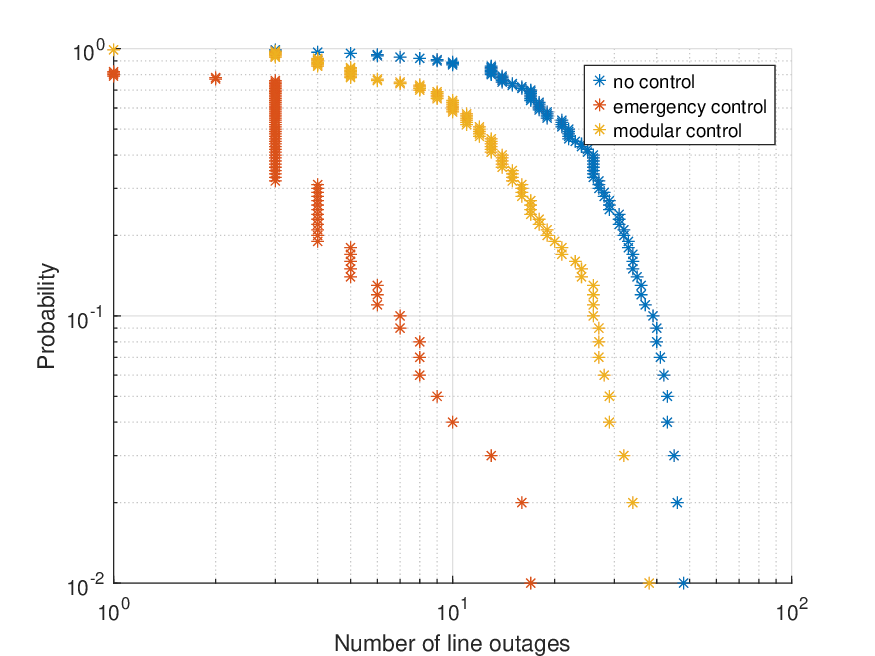}
	\vspace{-20pt}
	\caption{Complementary cumulative distribution (CCD) plot of the line outages for the IEEE 300-bus system under three scenarios.}
	\label{log_line_outages}
\end{figure}

The complementary cumulative distribution (CCD), i.e., log-log plot, of the blackout size in terms of MW lost and its occurrence of the Monte Carlo simulations is shown in Fig. \ref{loglog_plot_300}. The system without control shows power-law behavior. It can also be observed from the figure that blackout size reduces when using the modular control and reduces even more when emergency control is applied to the system under the same condition, which aligns with our previous analysis. Similarly, Fig. \ref{log_line_outages} shows a log-log plot of the line outage probabilities for the IEEE 300-bus system, which also verifies our approach compared to several other approaches adopted in the literature \cite{kaisun}.

\begin{figure} [h!]
	\centering
	\includegraphics[keepaspectratio=true,angle=0, width= 1 \linewidth]{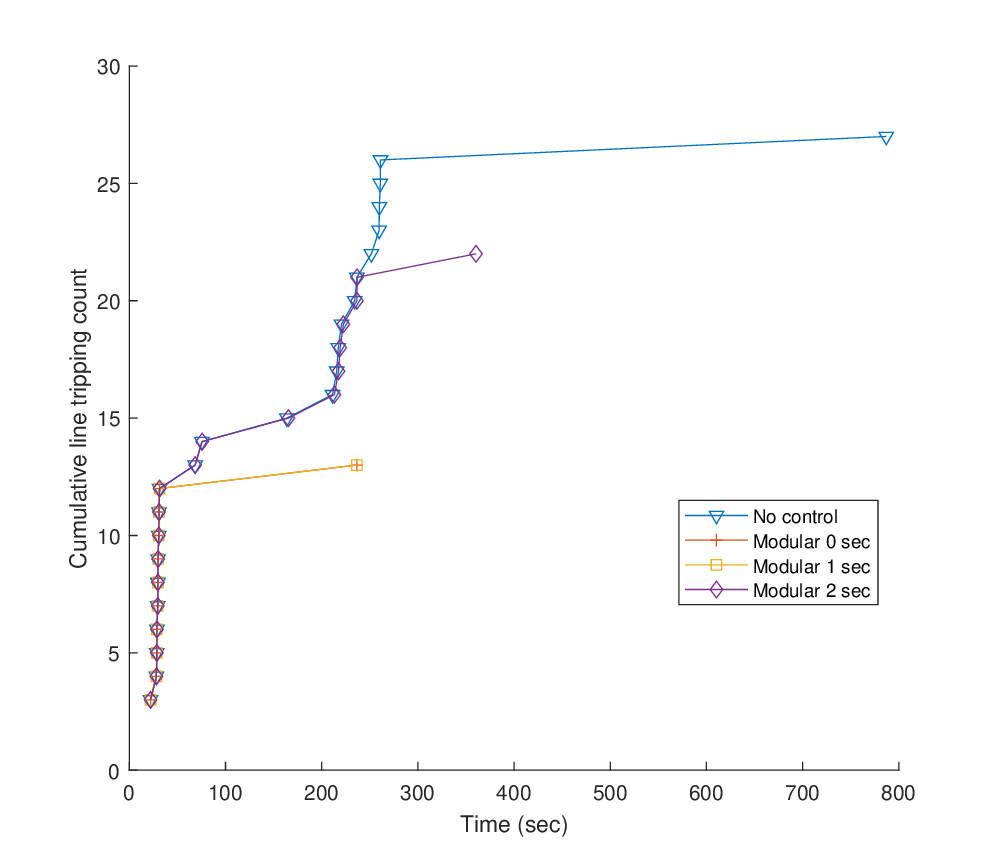}
	\vspace{-20pt}
	\caption{Impact of communication delay on the proposed modular control performance.}
	\label{Comm_delay}
\end{figure}

Additionally, we investigated the effect of communication delays between modular controllers. Fig. \ref{Comm_delay} shows a specific scenario for the IEEE 300-bus system, pair lines number 196 and 125 were tripped simultaneously as the initial $N-2$ trip. Three cases were simulated and compared against the case not using any control: 1) The proposed DES modular control without any communication delay, 2) DES modular control with a 1-second delay, and 3) DES modular control with a 2-second delay. Results show that as the communication delay for each modular controller increases, the response of the controller is also delayed, which may eventually lead to an increase in losses during corrective control actions, and in some cases, the failure of the controller. Under a centralized control, the control delay is longer than in modular control, which can lead to a worse result and a cascading failure.

The proposed approach was compared with other methods, as shown in Table \ref{Table_5}, in terms of \textquotedblleft control losses\textquotedblright due to load shedding and generation re-dispatch, and in terms of \textquotedblleft load lost\textquotedblright due to subsequent outages. In the table, for emergency control and the modular DES control, we selected specific $N-2$ cases, the same cases shown in Fig.  \ref{118_bus_single_case} and Fig. \ref{300_Bus_specific_case}. For the Markovian tree method, a specific $N-3$ initial outage was applied to the RTS 96-bus system \cite{kaisun}, \cite{ManagementofCascading}. We converted the economic-based metrics to MW-based losses. Although a different case study was used, the results align with our earlier analysis: under centralized control methods, losses are generally less under ideal conditions, with no communication delays and full observation by the centralized controller.	

\begin{table*}[h!]

	\caption{Comparison of results among different methods}
	\vspace{-20pt}
	\label{Table_5}
	\footnotesize
	\begin{center}
		\begin{tabular}{|l l l l l l|} 
			
			\hline
			Method & Control losses & Load lost &Scalability & Test case & Remarks \\ [0.5ex]
			
			\hline\hline
			Modular DES & 235.8 MW & 2171.4 MW & Decentralized & IEEE 300-bus system & These result apply to a specific $N-2$ initial outage.\\
			& 113.3940 & 0 & & IEEE 118-bus system & Topology and MW ratings widely differ between case studies\\
			\hline
			
			Emergency control &	182.05  & 1638.9 MW & Centralized   & IEEE 300-bus system & \\
			& 13.4 MW & 0 &  & IEEE 118-bus system & \\
			\hline
			Markovian tree & 5.59 MW &  1.3 MW & Centralized
			& RTS-96 & Topology and MW ratings widely differ between case studies.\\
			&  &   &  &  &  These result apply to a specific $N-3$ initial outage.\\
			\hline
		\end{tabular}
		
	\end{center}
\end{table*}

\section{Conclusion}

In this paper, we developed a new control to prevent cascading failures in large scale power systems using modular supervisory control of discrete event systems (DES). The modular control reduces computational complexity and increases the effectiveness and  robustness of the controlled system. 
We first extended the modular control approach to include forcible events and expanded the specification language to incorporate more events from neighboring nodes to improve control actions for each modular controller. 
We proposed a platform based on MATLAB environment to implement the modular strategy that couples the DES and continuous-time parts of the control and tested them with power system simulations. 
To verify the effectiveness of the proposed approach, we conducted a case study using the IEEE 30-bus, 118-bus and 300-bus systems. The proposed modular supervisory control can successfully stop cascading failures and significantly reduce the MW lost due to the failures. 
However, compared with the centralized control approach that we proposed in \cite{LLPWasseem}, the modular control approach results in more MW lost, and more lines tripped after observing an N-2 contingency. 
Nevertheless, the modular supervisory control is more reliable since it does not depend on a central controller, and each node only requires information from its neighbors. Our simulation results demonstrate the effectiveness of the proposed approach in mitigating cascading failures in power systems.



\bibliographystyle{ieeetr}

\bibliography{PowerS}

\end{document}